\documentclass[aps,pre,preprint,superscriptaddress,longbibliography]{revtex4-1}

\usepackage{amsmath,amssymb,pifont}
\usepackage{graphicx,color,subfigure,bm}
\usepackage[ansinew]{inputenc}

\def\etal{\emph{et al. }}

\begin{document}


\title{Some fluid mechanical aspects of artistic painting}


\author{Roberto Zenit}
\email{zenit@unam.mx; zenit@brown.edu}
\affiliation{Instituto de Investigaciones en Materiales, Universidad Nacional Aut\'onoma de M\'exico, Apdo. Postal 70-360, Coyoacan, Ciudad de Mexico 04510, Mexico}
\affiliation{School of Engineering, Brown University, 184 Hope St, Providence, RI 02912, USA}


\date{\today}

\begin{abstract}
Painting is a fluid mechanical process. The action of covering a
solid surface with a layer of a viscous fluid is one of the most
common human activities; virtually all man-made surfaces are painted to
provide protection against the environment or simply for
decoration. This process, in an industrial context, has been vastly studied and it is well
understood. In case of artistic painting the purpose is different. Painters
learn how to manipulate the non-uniform deposition of paint onto a surface, through lengthy empirical testing
of the action and modifying the physical properties of the fluids, to create
textures and patterns  of aesthetic value. In
this paper, an analysis of some notable  painting techniques is presented from the point of view of fluid mechanics. In particular, we discuss the so-called `accidental painting' technique, originally
devised by David A. Siqueiros, which is the result of a Rayleigh-Taylor
instability. An analysis of several techniques used by Jackson
Pollock is also presented, showing how he learned to carefully
control the motion of viscous filaments to create his famous
abstract compositions. We also briefly discuss how pattern and textures are produced in decalcomania and watercolor painting. These investigations indicate that it is possible to establish concrete scientific discussions among modern
fluid mechanics, art, art history and conservation.
\end{abstract}


\maketitle

\section{Introduction: painting is fluid mechanical}

One of the distinctions between humans and other animals is the fact that the former are capable of painting: depicting natural or imaginary images onto a piece of cloth, paper or any other surface simply for decoration. No other life-form, regardless of intelligence, does this. Art is therefore a unique human feature and painting is one of the oldest visual-art forms: the drawings at the \emph{Grotte Chauvet} in France are considered to be the first formal paintings, believed to be about 32,000 years old.

Paint is a fluid composed of a solvent and a solute that, applied on a surface and after a process where the solute evaporates, forms a solid coating composed of pigments. Painting, on the other hand, is a mechanical endeavour: a volume of fluid is placed onto a surface to then be  spread with the use of a tool to form a thin layer. The formation of such thin layer depends on the fluid physical properties (viscosity, density and surface tension), on the tool (shape and orientation) and on the action (speed of tool and separation distance from the substrate). Basically all man-made surfaces are painted either to improve their resistance to the environment or for decoration. To reduce economic losses,  industrial painting has proposed many strategies to guarantee uniform coatings. Nowadays this industry is a mature field \cite{Lambourne1999,Seiwert2013}, and the process to generate a uniform layer of paint as fast as possible is well understood \cite{Landau1942,Taylor1960,Higgins1979}.

Artistic painting is different. In most cases non-uniform coating is what actually is desired. The textures that result from paint non-uniformity are used as important parts of the composition. Artists learn how to create paint traces without any technical fluid-mechanical training, by pure empiricism, lengthy repetition and experimentation. Naturally, throughout history, the techniques used by artists have evolved; however, the appearance of modern materials, paints and substrates, have opened new and large avenues of creation. Due to its fundamental difference with industrial painting, the study and understanding of artistic painting has scarcely been addressed. The physical mechanisms that lead to non-uniform coatings remain unexplored.
We note, however, that the fluid mechanics community has had a sympathetic eye for aesthetics. The great popularity of the gallery of fluid motion is the perfect example of such predilection  \cite{Samimy2014,Herczynski2016,Tretheway2017,Sharp2018}.

In this paper, we report on a few recent efforts to formally study artistic painting as a modern fluid mechanics problem. Interestingly, at least in the cases presented, the formation of patterns is driven by well-studied hydrodynamic instabilities. In some instances, the artists purposely avoid the appearance of instabilities; in others, the instabilities are sought after. We make references and links to current studies in fluid mechanics that help explain how the patterns are created, or how they can be improved or controlled. We also point out how the study of a particular painting technique can lead to interesting new investigations in fluid mechanics. This paper is also a personal recount of the adventure to leave the comfort established research subjects. Hopefully these ideas inspire some readers to `look for the edges'.

The paper is organized as follows. We first will discuss the painting techniques that are associated with particular artists; in particular, we will show some examples of the works of David A. Siqueiros and Jackson Pollock. Then, we will focus on other painting techniques which are not necessarily associated with a particular painter. We close by recounting the experiences of collaborating with working artists and how these investigations have influenced their recent work.

\section{Artists}

\subsection{Siqueiros' accidental painting technique was not accidental}

David Alfaro Siqueiros  is one of the most notable Mexican painters of the XX$^{th}$ century \cite{folgarait1987}. As with most artists,  his painting technique and subject matter evolved during his life, but he is most well known for his murals filled with social themes. Early in his career he gained notoriety for his interest in finding new ways to paint. His proposal to experiment with new painting tools included the use of rapid-drying paint, sprays, horizontal canvases (often wood and other) and some other materials \cite{McGlinchey2013,zetina2013}. An example of the influence and leadership of Siqueiros, in his quest to find new techniques, was the fact that he organized an experimental painting workshop in New York in 1936 \cite{Hurlburt1976,Hurlburt1976b}. In it, many artists gathered to learn and experiment.

\begin{figure}[ht!]
    \centering
    \subfigure[]{\includegraphics[width=0.45\textwidth]{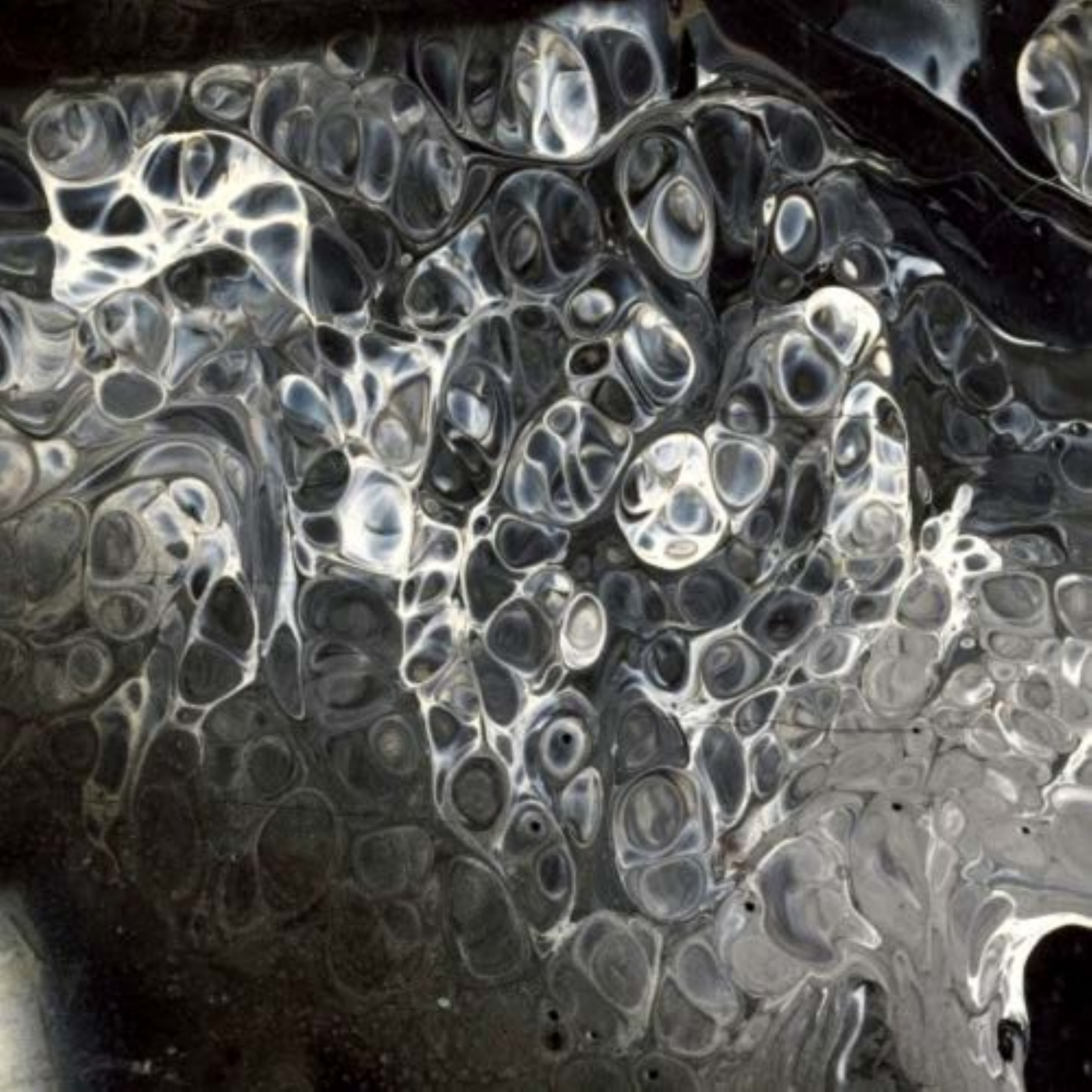}}  \,
    \subfigure[]{\includegraphics[width=0.45\textwidth]{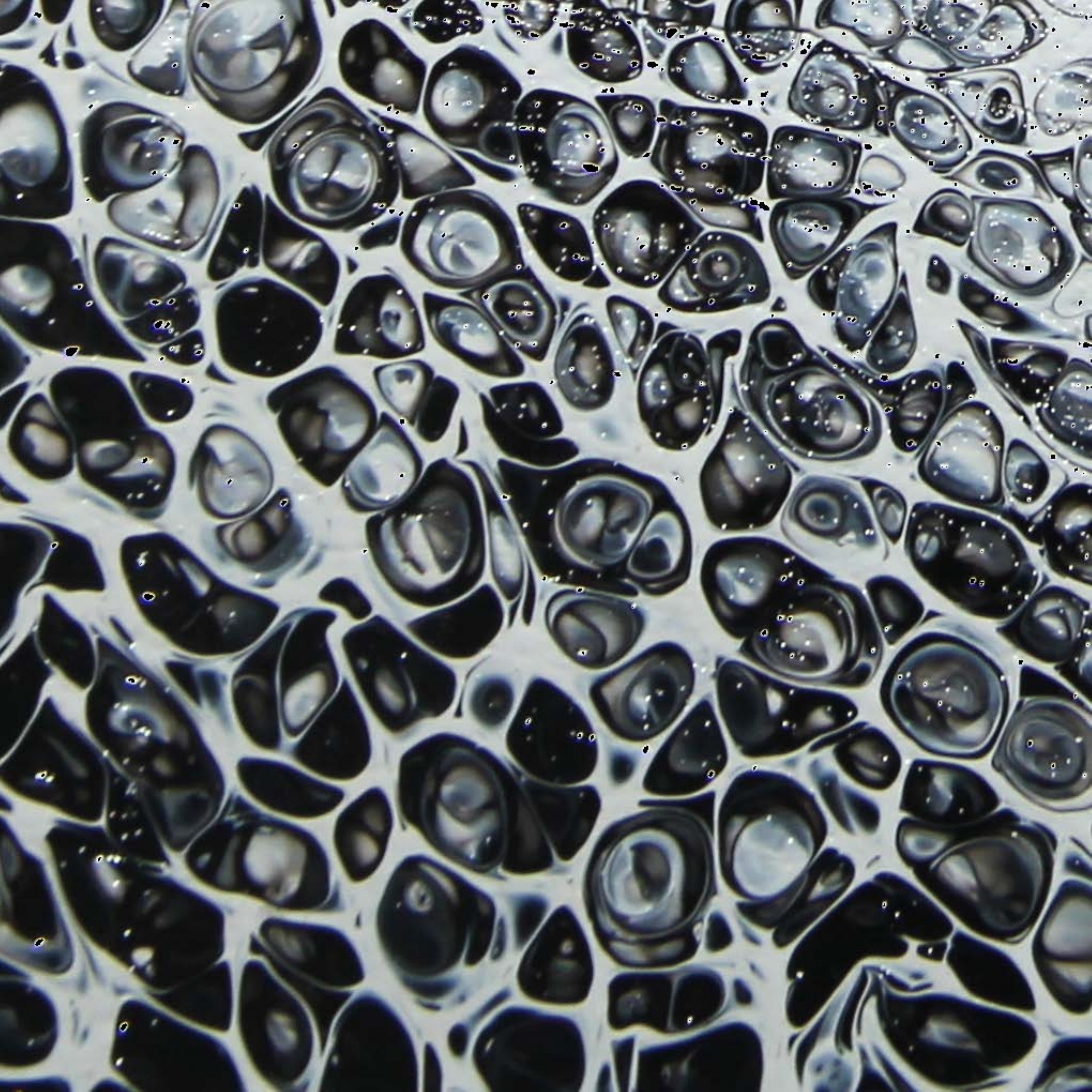}}
    \caption{(a) Zoom of `Collective Suicide' D.A. Siqueiros, 1936, Museum of Modern Art, New York. Photograph by A. Aviram, reproduced with permission. (b) Pattern created under controlled conditions, following \cite{Zetina2015}. Both images are approximately the same size.}
    \label{fig:siqueiros1}
\end{figure}

Among the painting techniques that were `invented' during this workshop, we became interested in the accidental painting technique. The technique leads to the formation of well-defined characteristic patterns that appear in several paintings by Siqueiros from that time. Figure \ref{fig:siqueiros1}(a) shows a typical example of such patterns. Siqueiros himself loosely, and poetically, described how the `method of absorption' gave rise to the patterns \cite{Siqueiros1936}. In essence, Siqueiros would pour layers of paint of different colors on top of each other over a horizontal substrate. The patterns would naturally emerge without the action of the painter after some time. 

In \cite{Zetina2015} we studied the technique in a controlled manner, following the description in \cite{Siqueiros1936}. What we found was that the patterns were the result of the instability of the fluid layer. For the cases when the top layer was denser than the lower one, the configuration became Rayleigh-Taylor unstable \cite{rayleigh1883,taylor1950}. Figure \ref{fig:siqueiros1}(b) shows an example of the patterns created in the laboratory which hold striking similarities to those produced by Siqueiros. This instability has been extensively studied in our community in a wide variety of contexts, ranging from oceanography, atmospheric science, combustion and even astronomy.  In our study we made comparisons of the size of the cells observed in the experiments and the predictions from a linear instability calculation, leading to good agreement. For details about the experiment and calculation, please refer to \cite{Zetina2015}.

\begin{figure}[ht!]
    \centering
    \subfigure[\,high viscosity paint]{\includegraphics[width=0.45\textwidth]{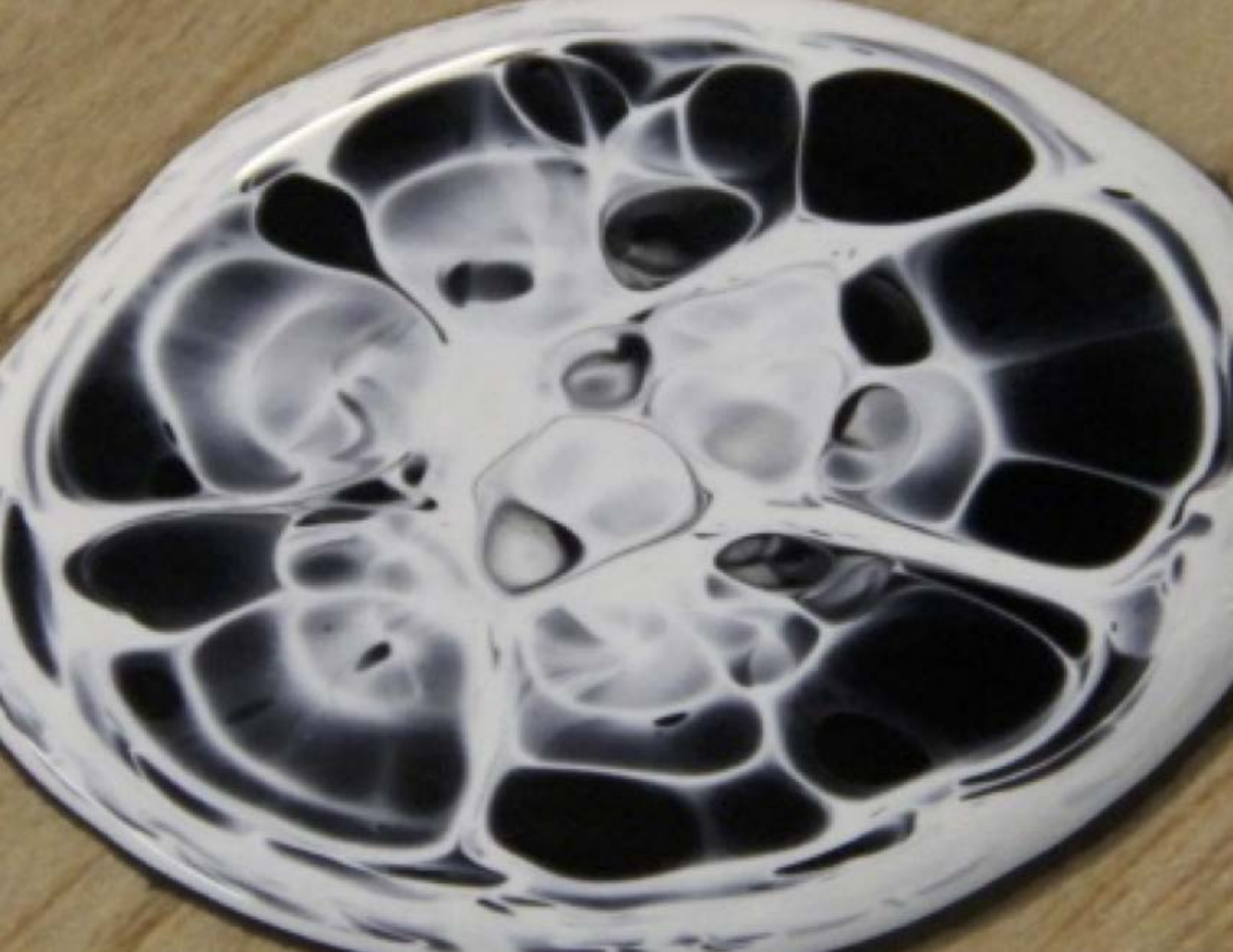}}  \,
    \subfigure[\,low viscosity paint]{\includegraphics[width=0.45\textwidth]{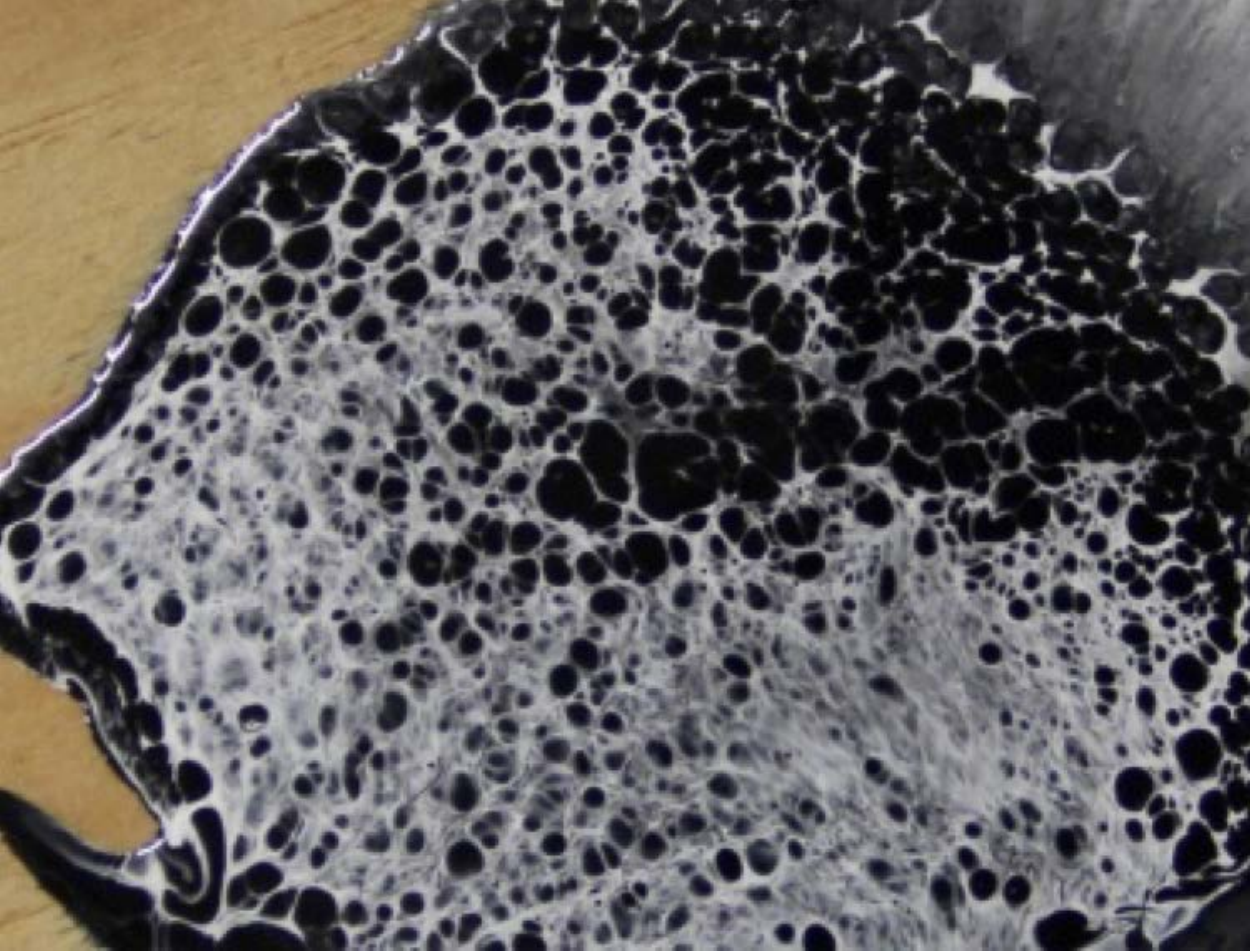}}
    \caption{Experiments showing the change in size of the patterns by changing the paint viscosity. The visual observations are in qualitative agreement with the linear instability calculations from \cite{Zetina2015}. Both images are approximately the same size.}
    \label{fig:siqueiros2}
\end{figure}

Since the nature of process for pattern formation is  understood, it is possible to play with the parameters space to `design' the size of the instabilities. By adding thinner to the paints, the viscosity changes drastically, while the density remains relatively unchanged. According to the linear instability calculations in \cite{Zetina2015}, decreasing the viscosity leads to a faster  growth rate of the instability but does not affect significantly the size of the most unstable mode. In turn, reducing the viscosity causes the thickness of the layers to be thinner, which leads to a decrease of the instability cell size. Figure \ref{fig:siqueiros2} shows images of a series experiments conducted for fluid layers with different fluid viscosities. In both cases, the experiment was conducted considering the same nominal conditions (same volume of the two paints). Although neither the change in viscosity nor the size of the cells were quantified, the images clearly show the differences and show a good  qualitative agreement with the linear instability calculation.

Discovering the physical foundation behind this pattern and texture formation was fascinating. In addition to learning about the life of Siqueiros and his incredible artistic vision, we realized that many aspects of the painting technique held surprising similarities to flow phenomena in other areas. In particular, we found some similarities with the formation of lava domes \cite{mckenzie1992}. Figure \ref{fig:siqueiros3}(a) shows an image of a pancake-like dome photographed in Venus. We can conjecture that the pattern observed on the surface of the domes may be the result of a similar physical mechanism as that discovered for the accidental painting technique. The temperature gradient, from the base to the free surface of the dome, would induce a density gradient that would lead to {an unstable layer}. As in the case of paint, the lava would eventually solidify capturing the patterns created by the instability. {Please note that when the instability is driven by a temperature gradient, it is called Rayleigh-Bernard instability\cite{Bodenschatz2000}. Both instabilities share many features but are different in nature. Another system of interest is the internal structure of the salt diapirs in the Great Kavir desert in Iran, shown in Fig. \ref{fig:siqueiros3}(b). The observed salt upwellings may be the result of a similar unstable mechanism \cite{jackson1990}. In that case, the density gradients result from differences in evaporation rate. Although we have not attempted a formal quantitative analysis of these flows, in the context of the Rayleigh-Taylor instability, the similarities are noteworthy. 

\begin{figure}[ht!]
    \centering
    \subfigure[]{\includegraphics[width=0.45\textwidth]{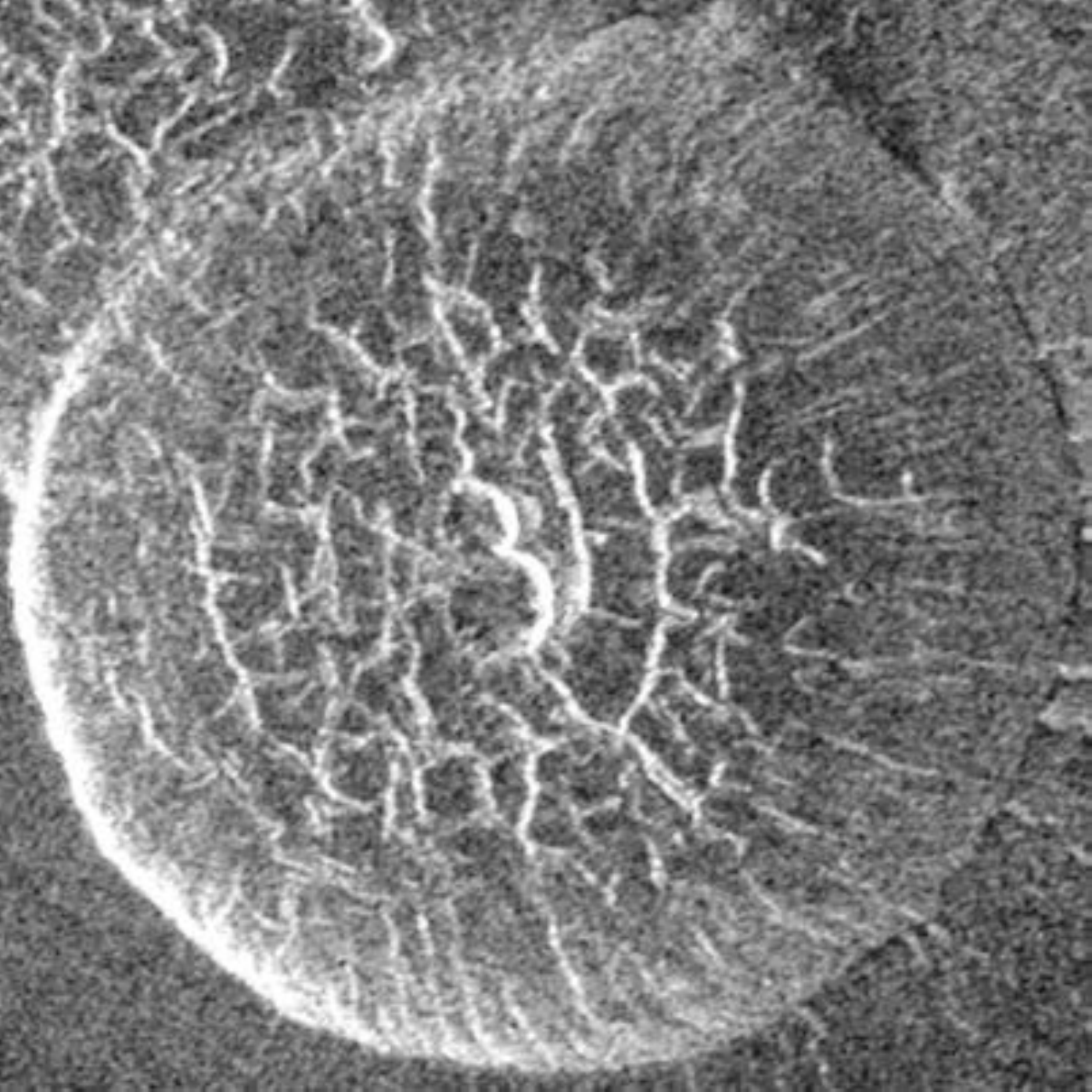}}  \,
    \subfigure[]{\includegraphics[width=0.45\textwidth]{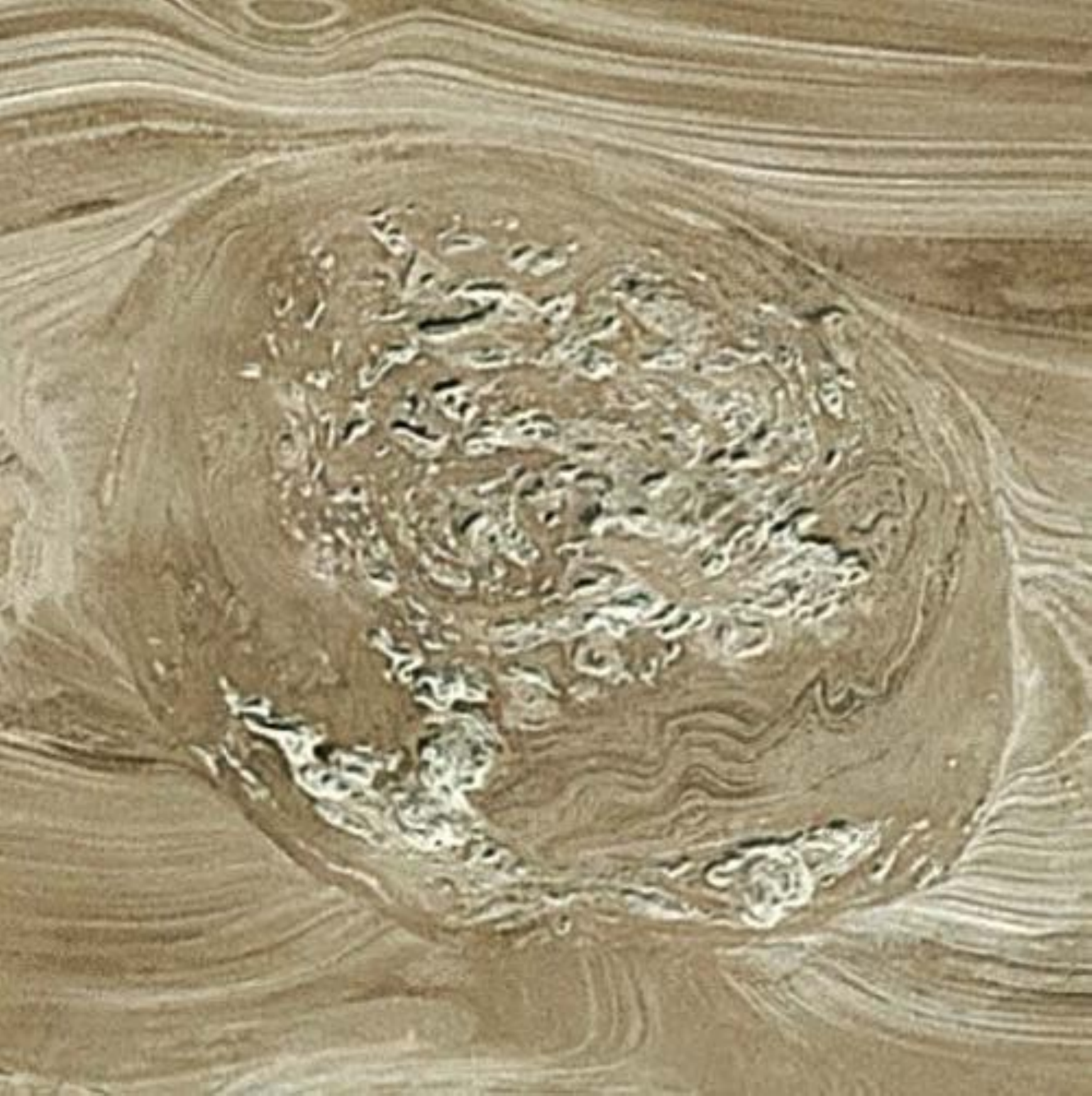}}
    \caption{Examples of two flows with the same physics as the accidental painting technique. (a) Pancake lava dome in Venus. Image from Magellean mission, NASA/JPL. (b) Salt diapir from of the Great Kavir desert in Central Iran. Image from Landsat 7 imager, NASA/USGS.}
    \label{fig:siqueiros3}
\end{figure}

\subsection{Pollock and fluid filaments}

Jackson Pollock is, perhaps, the most famous American painter of last century. The  paintings from his so-called `dripping' period (ranging from 1947 to 1950) are easily recognized by the webs of paint lines and splashes that created abstract textures that have fascinated general audiences and experts alike. An example is shown in Fig. \ref{fig:pollock1}(a).

One distinctive feature of Pollock and his painting technique is that it is well documented. In addition to writing about it himself \cite{pollock1947}, he allowed to be photographed \cite{Namuth1980} and filmed while working \cite{Namuth1950}. These films grant unprecedented access to his painting technique. Figure \ref{fig:pollock1}(b) shows an image of Pollock, photographed by Hans Namuth, in his studio while painting. The technique, incorrectly named `dripping', consists of depositing viscous fluid filaments while moving the hand around a horizontal canvas and at certain height. The paintings produced in this manner are composed of fluid lines that are relatively straight. Interestingly, the origins of the dripping technique can be traced back to Siqueiros' experimental painting workshop; Pollock was among the attendees \cite{Hurlburt1976b}. 

\begin{figure}[ht!]
    \centering
    \subfigure[]{\includegraphics[width=0.6\textwidth]{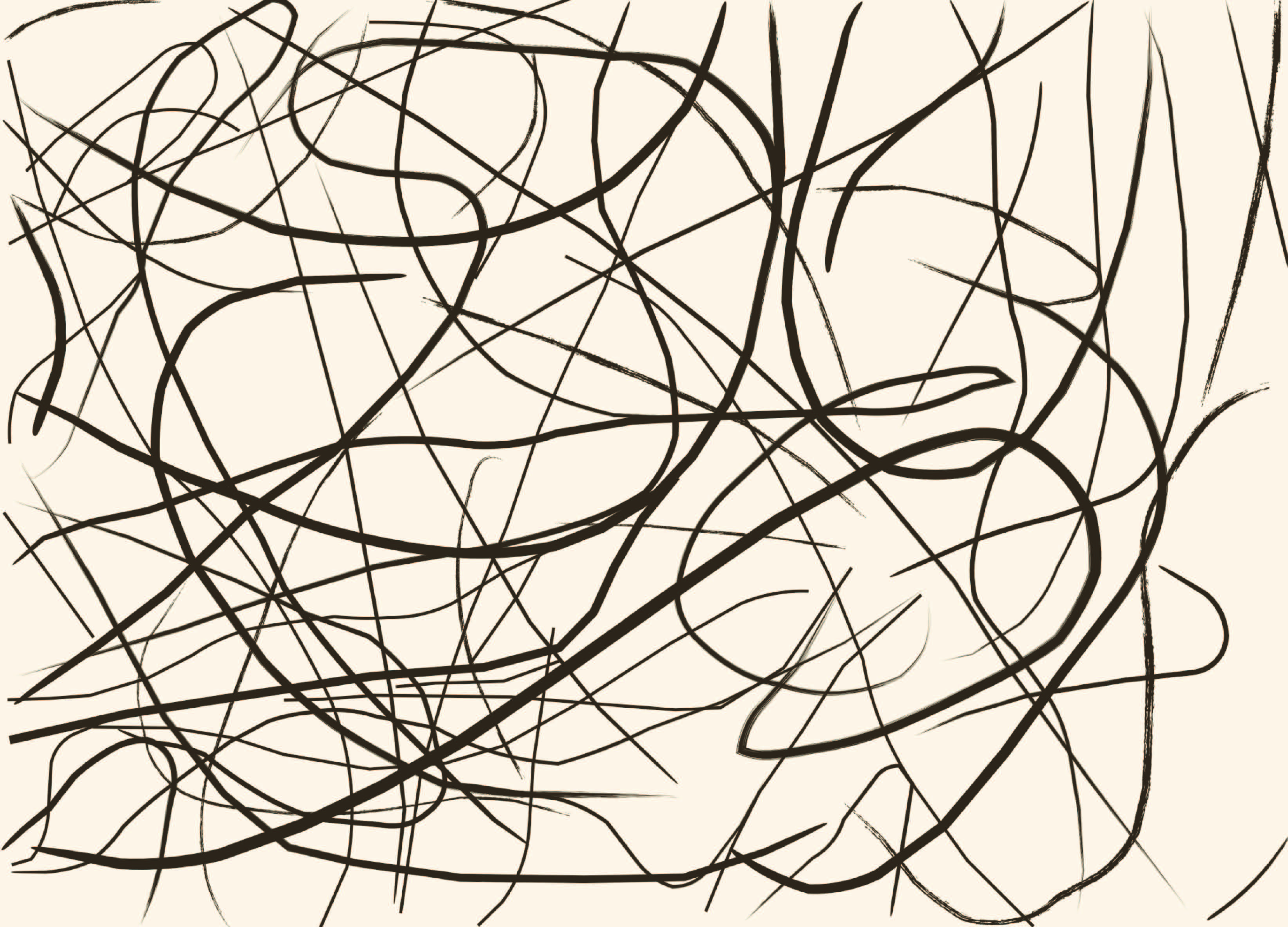}}\\
    \subfigure[]{\includegraphics[width=0.6\textwidth]{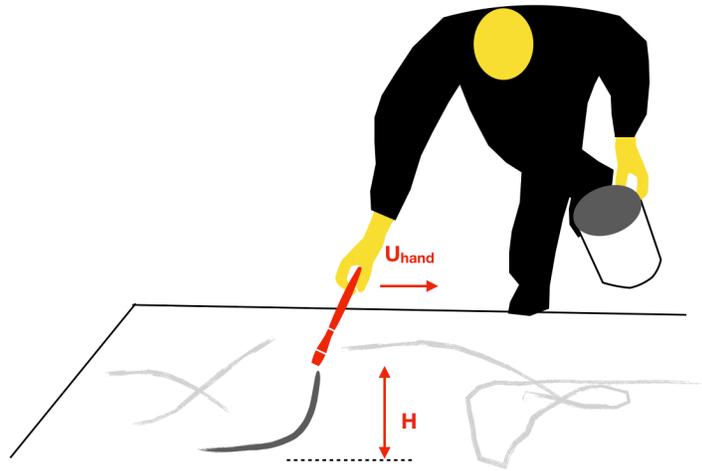}}    
\caption{(a) Sketch of the traces produced by Pollock. Copied from `Number 14 (Gray)', J. Pollock, 1948. The original image can be found at the Yale University Art Gallery web site. (b) Schematic view of Pollock painting, as in the film by Hans Namuth\cite{Namuth1950}. }
\label{fig:pollock1}
\end{figure}

Herczynski and collaborators conducted a seminal analysis of Pollock painting technique \cite{Herczynski2011}. In it, they discussed how  Pollock poured the liquid paint from the can (or using sticks or brushes) onto the canvas to form fluid filaments to draw with them, while moving around the canvas. They argued that some features of the painting were produced by the so-called coiling instability: the fluid filaments curl upon reaching the surface under the action of their own weight due to a viscous-buckling effect. This instability has been widely studied and it is relatively well understood \cite{Ribe2012}. The difference between this classical flow and the painting techniques used by Pollock is that in the former, the hand (or the painting instrument) moves as the fluid issues down. Therefore, the fluid filament is stretched out as it is being deposited on the canvas. So, a coiling filament is deposited along a horizontal path instead of onto itself. Chiu-Webster and Lister \cite{Chiu2006} coined the term `fluid mechanical sewing machine' to describe this phenomenon. The shape of the patterns left by this flow have also been well studied \cite{Morris2008,Welch2012,Brun2012,Brun2015}.

Herczynski \etal \cite{Herczynski2011} argued that the fluid filaments would cease to coil if the speed of the filament at the canvas surface was larger than the tangential coiling speed. 
Recently, we conducted a series of controlled experiments and video analysis of the historical Pollock videos \cite{Palacios2019}. Both hand speed, $U_{hand}$, and dropping height, $H$, were measured; also, the paint properties ($\mu$ and $\rho$, viscosity and density respectively) were estimated. With this information, we concluded  that, contrary to what Herczynski \etal proposed,  Pollock mostly painted under the conditions for which  the coiling instability would not occur. That is, Pollock move his hand sufficiently fast and poured paint from heights that would primarily lead to non-curled fluid filaments. Figure \ref{fig:pollock2} shows two examples of filaments below and above the transition from coiling to straight lines.
\begin{figure}[ht!]
    \centering
    \subfigure[\,$U_{hand}<\Omega R$]{\includegraphics[width=0.9\textwidth]{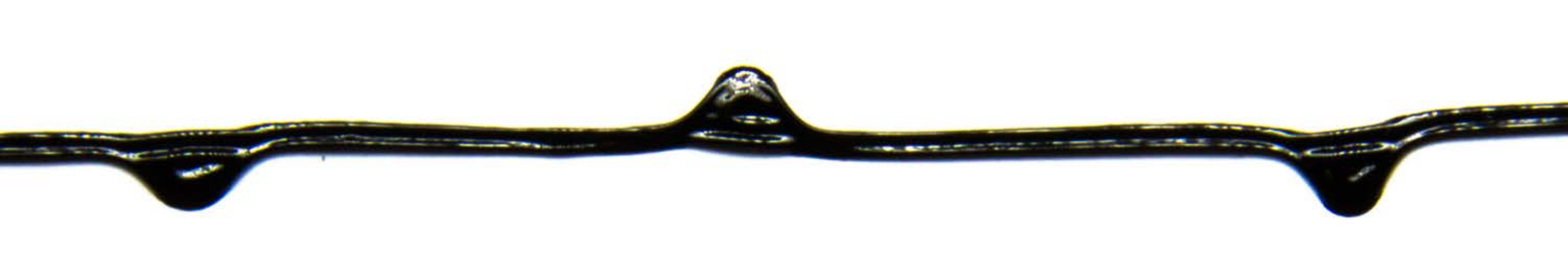}}  \\
    \subfigure[\,$U_{hand}>\Omega R$]{\includegraphics[width=0.9\textwidth]{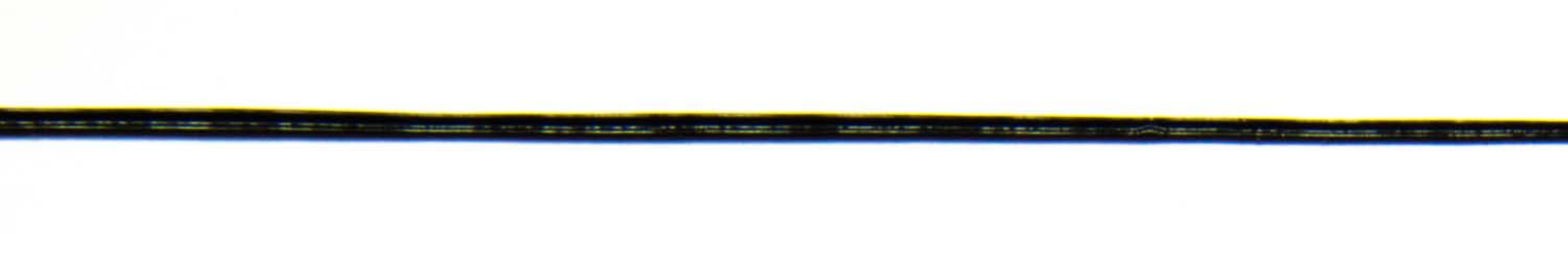}}
    \caption{Example of laid paint filaments before and after the transition from coiling to straight lines. Both dropped from the height $H=10$ cm, same flow rate and same viscosity, but different hand speed $U_{hand}$. $\Omega$ and $R$ are the coiling frequency and radius, respectively. Details of the experiment can be found in \cite{Palacios2019}.}
    \label{fig:pollock2}
\end{figure}

In addition to mostly avoiding the coiling instability, Pollock also painted with  paints that were sufficiently viscous such that the Rayleigh-Plateau instability would not occur. Therefore, his fluid filaments would rarely fragment. Based on the measurements of the the height and filament diameter, $d$, the slenderness, $H/d$, was around 300. Considering the estimation of paint properties, the Ohnesorge number, $Oh  = \mu/\sqrt{\rho \sigma d}$, was around 20. Even for such slender filaments, according to  \cite{Driessen2013}, fragmentation would not occur. As discussed by Eggers and Villermaux \cite{Eggers2008}  such viscous filaments are surprisingly stable. They appear to thin unboundedly. {Note that, however, single drops can be observed in many of Pollock's paintings, readily observable in Fig. \ref{fig:pollock2}(a). Since such drops do not appear to be deposited in an organized manner, we can argue that such `accidents' occurred during the loading process of the brush and were not purposely placed on the canvas.

Highly stable thin fluid filaments, such as those discussed above, have been observed in some Hawaiian volcanic eruptions. Known as Pele's hair \cite{Shimozuru1994}, the volcanic glass filaments, which are air-drawn from molten lava, can have a slenderness as high as 4000 without fragmenting for Ohnesorge numbers of about 150, considering the physical properties of of molten lava. The condition for such long viscous filaments to remain stable before solidifying were discussed by Villermaux \cite{Villermaux2012}, but the precise process that leads to their formation remains unexplained and mysterious.

\subsubsection{Flying viscous catenaries}
Early in his career, before fully engaging in the dripping technique, Pollock mostly painted with brushes. However, it is believed that he kept on experimenting with other techniques resulting from Siqueiros' influence \cite{Hurlburt1976b}. Recently, the Getty Conservation Institute undertook an extensive conservation  project of `Mural', which is a large painting from 1943 (before the dripping period) which marks in important transition point in the development of Pollock's professional career \cite{Szafran2014}. In addition to the conservation procedure of the piece, among the many studies that were conducted, one particular feature sparked our interest. In many parts of the painting, distinct paint filaments were observed. In Fig. \ref{fig:pollock3}(a) such features are clearly shown (pink paint). These viscous filaments could not have been deposited with the dripping technique since the piece was executed with canvas oriented in the usual manner (standing, as opposed to horizontal). 
Researchers at the Getty Conservation Institute  conjectured that the curly lines resulted from flicking a brush filled with viscous paint \cite{Getty2014}, as depicted in Fig. \ref{fig:pollock3}(b). The paint would be expelled from the brush, forming stable flying viscous filaments and that would eventually reach the canvas. 

\begin{figure}[ht!]
    \centering
    \subfigure[]{\includegraphics[width=0.6\textwidth]{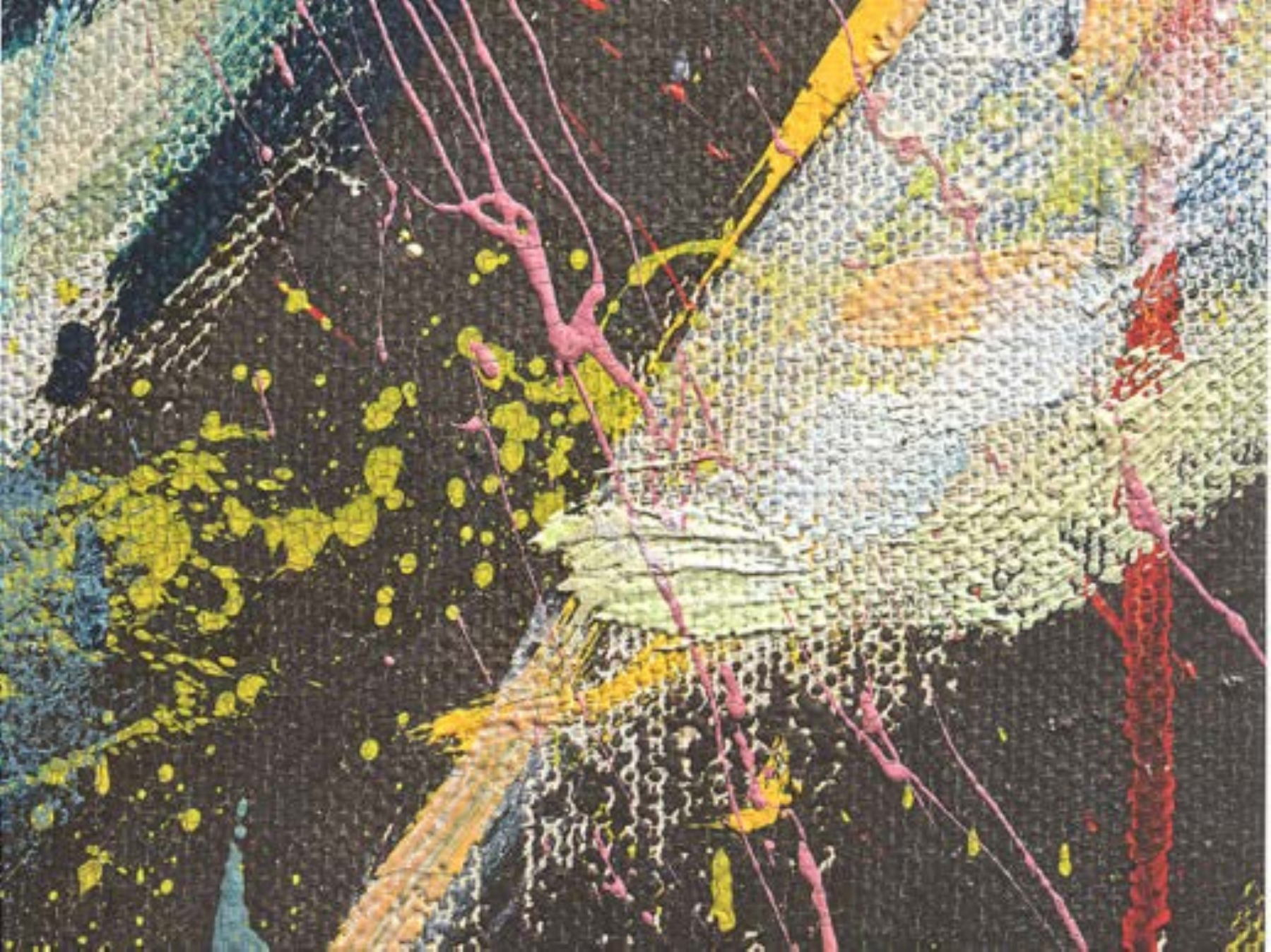}}  \\
    \subfigure[]{\includegraphics[width=0.6\textwidth]{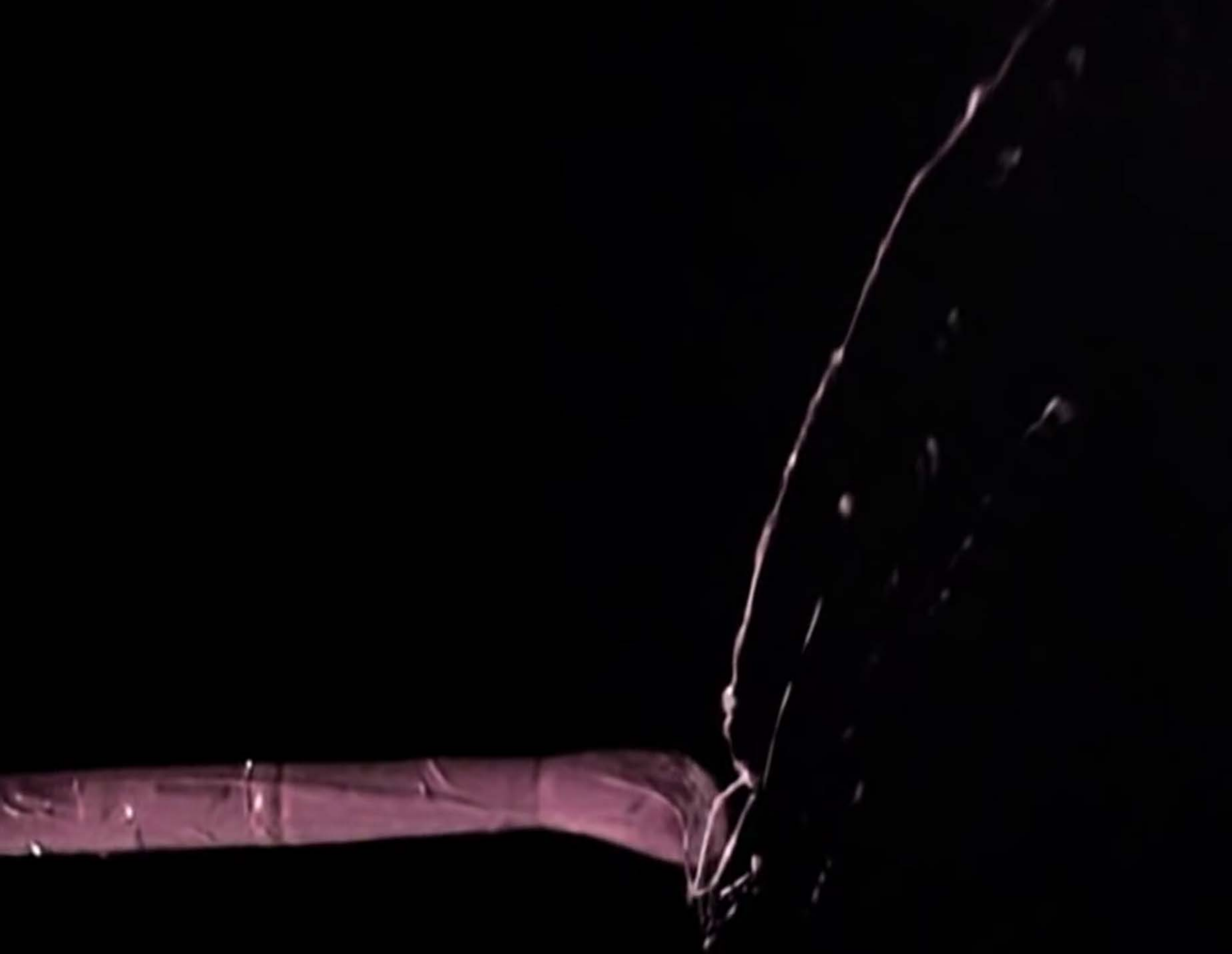}}
    \caption{(a) Zoom of `Mural', J. Pollock, 1943, University of Iowa, Stanley Museum of Art. Reproduced with permission.(b) A paint-filled brush being `flicked' to expel liquid filaments. Image taken from \cite{Getty2014}, used with permission.}
    \label{fig:pollock3}
\end{figure}

To study the behavior of these flying viscous stable catenaries in a more controlled fashion, we conducted a few simple experiments. A paint-filled brush was mounted on a rotating arm, which was rotated at a constant rate and, by changing the length of the arm, it imposed a constant tangential speed to the tip brush, $U_{tip}=\omega L$. Figure \ref{fig:pollock4} shows side views of the filaments that were formed by varying the rotation rate and the paint viscosity. In addition to the Ohnesorge number, we also define a a dimensionless brush acceleration $G=U_{tip}^2/(gL)$. As expected, the fluid could not be expelled from the brush if $G<1$ (Fig. \ref{fig:pollock4}.a), but flying catenaries that were useful to `paint' were observed for $G>10$. Also, stable unbroken filaments were produced for $Oh>10$, as shown in Fig. \ref{fig:pollock4}.(b). A few experiments were also conducted with paints with reduced viscosity. The filaments would form but fragment in mid air, as shown in Fig. \ref{fig:pollock4}.(c).
\begin{figure}
    \centering
    \subfigure[\,$G=0.8, Oh=17l7$]{\includegraphics[width=0.32\textwidth]{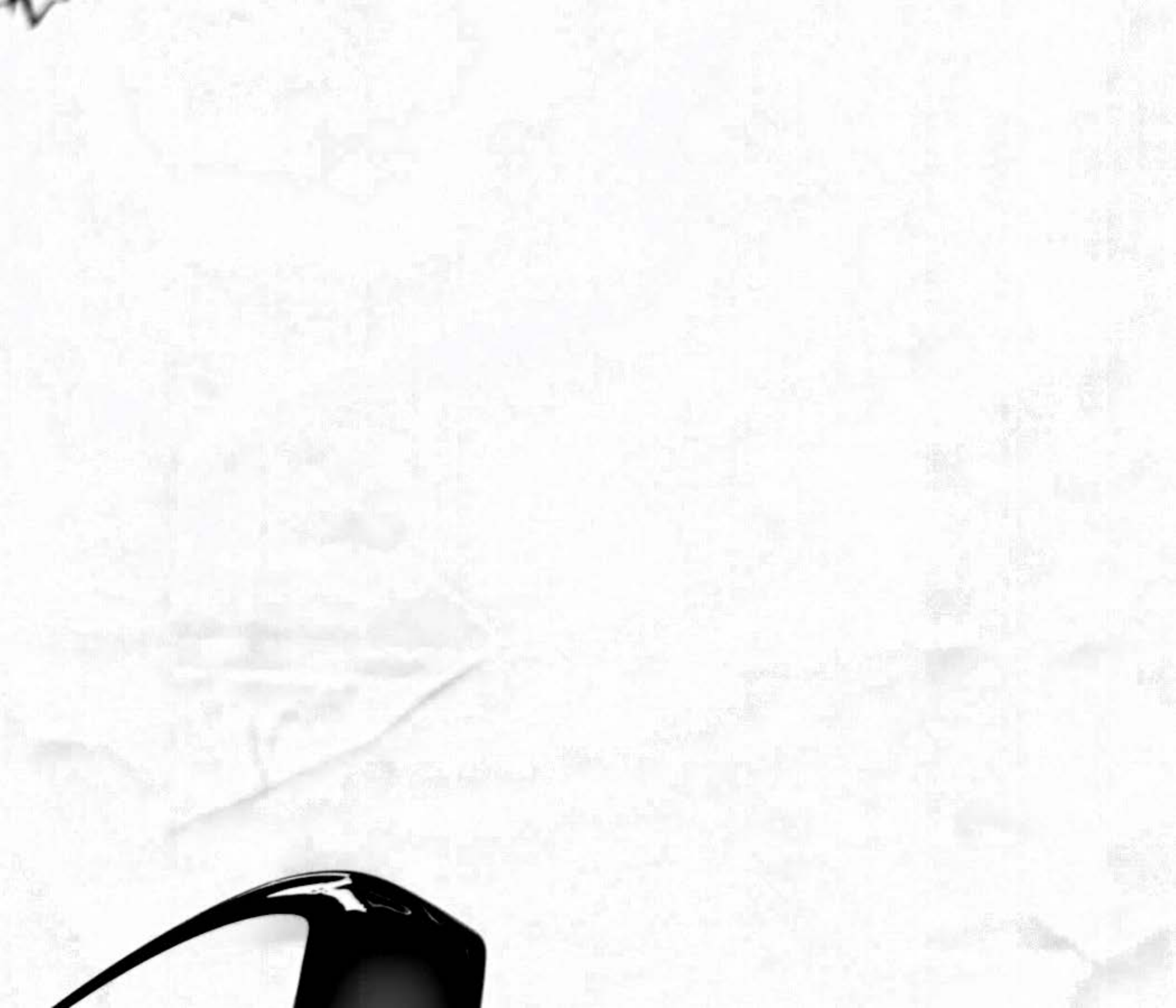}}  \,
    \subfigure[\,$G=22.2, Oh=17.7$]{\includegraphics[width=0.32\textwidth]{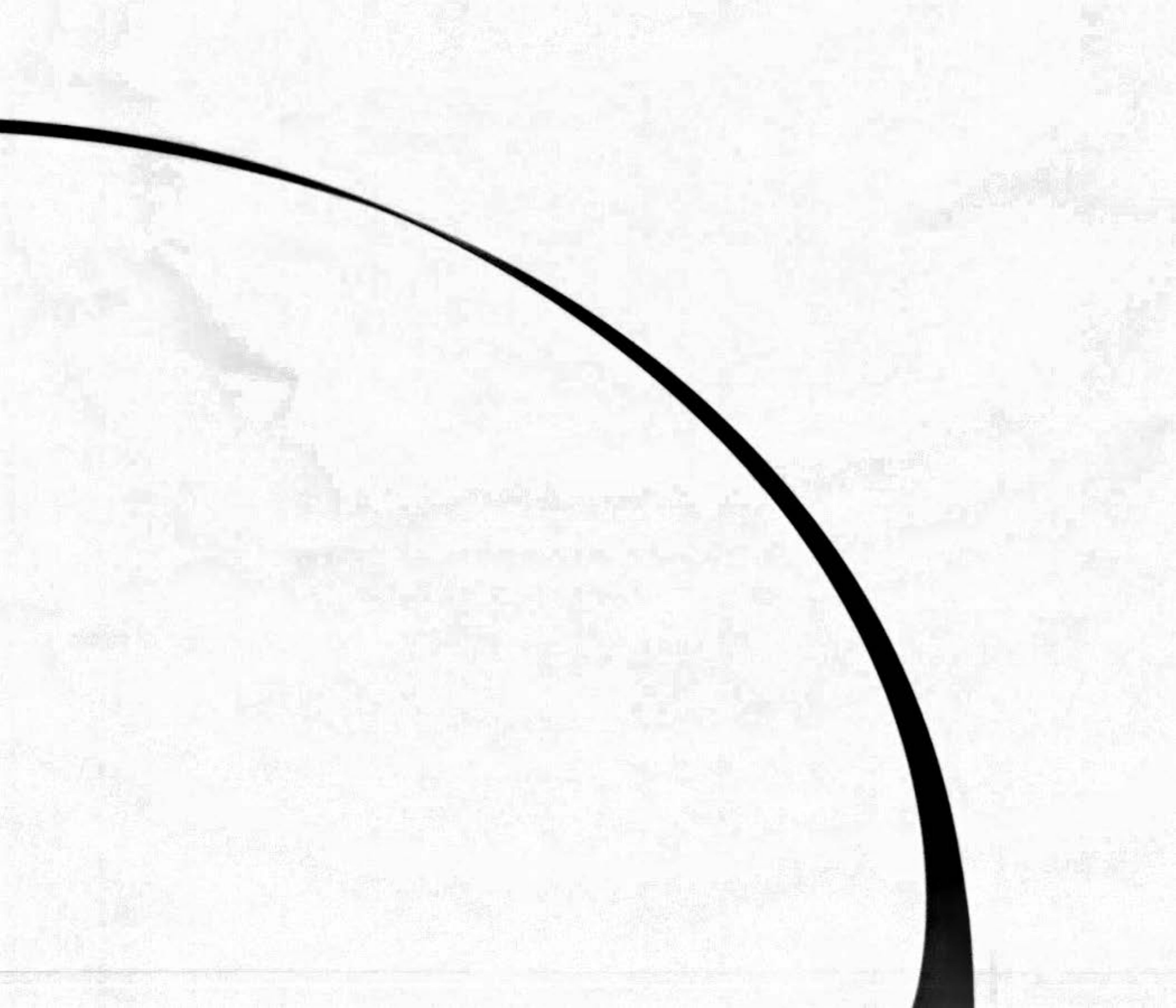}}  \,
    \subfigure[\,$G=22.2, Oh=0.02$]{\includegraphics[width=0.32\textwidth]{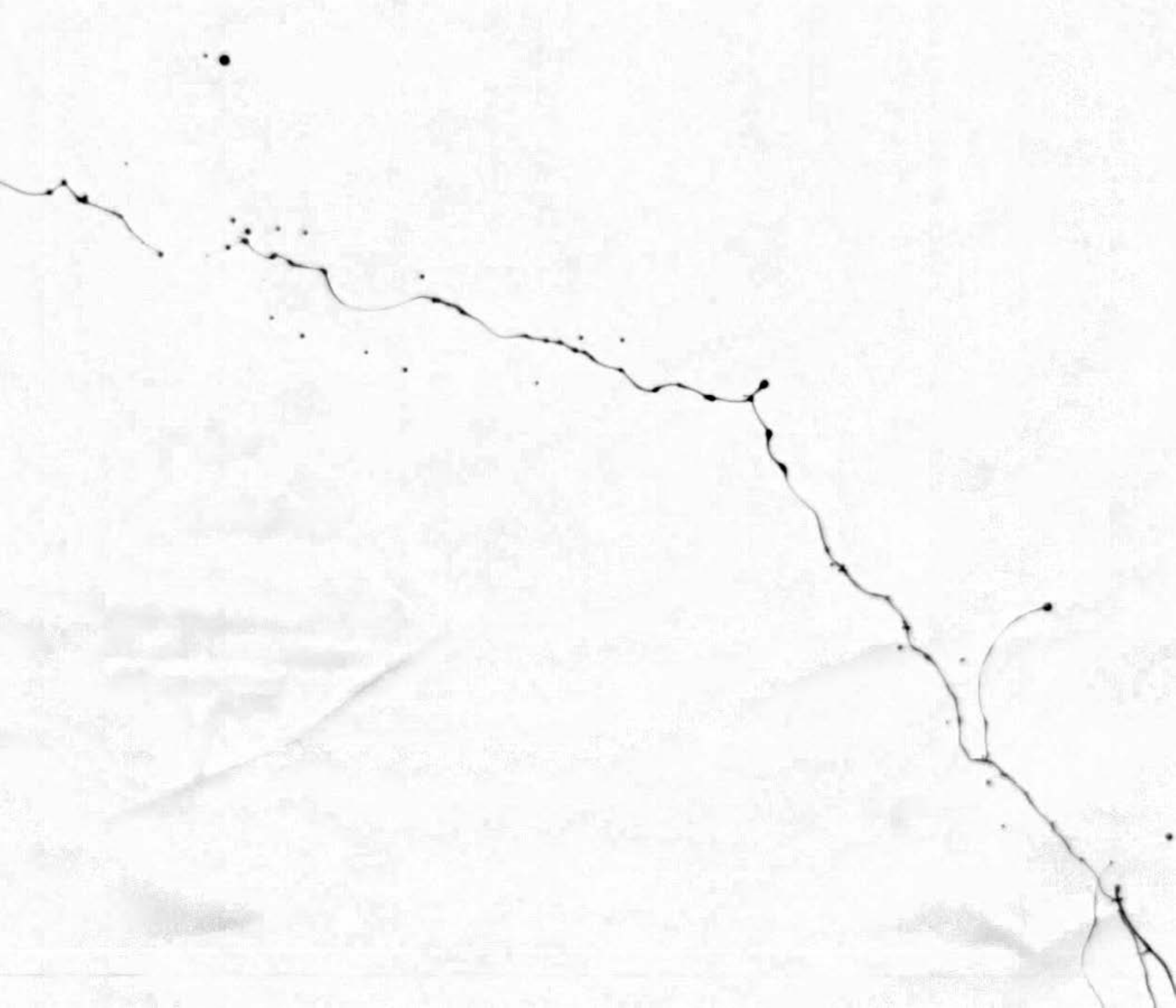}}  \\
    \subfigure[\,$G=22.2, Oh=17.7$ (landing on canvas)]{\includegraphics[width=0.9\textwidth]{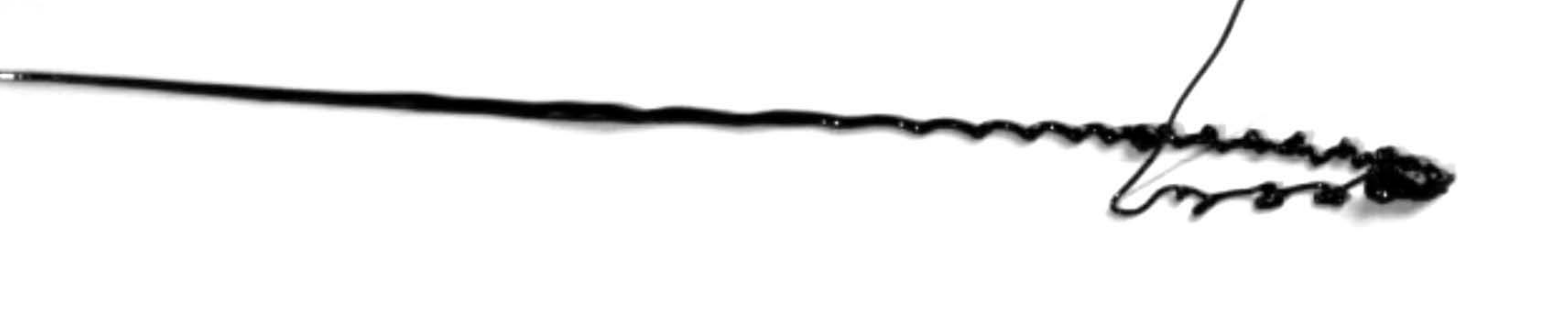}}
    \caption{Side view of flying viscous catenaries. (a) for small $G$ the catenary does not form. (b) For large $G$ and $Oh$ a stable catenary forms. (c) For small $Oh$, the catenary forms but fragments in mid-air. (d) The trace forming while the viscous catenary lands on a vertical canvas. The image was rotated 90$^o$, clock-wise.}
    \label{fig:pollock4}
\end{figure}

Once the catenaries were successfully deployed, they could land on the surface of a vertically oriented canvas, some distance away. Depending on the distance, the filaments could reach the surface at different angles. When the angle was small, the filament would form a straight line; on the other hand, at large angles, the filament would coil in a manner similar to that for liquid coiling \cite{Ribe2016}. Additionally, some waviness in the final drawing was also the result of the interaction of the filament with the air, reaching air Reynolds numbers of O(10$^3$). An example of the trace formed by a landing catenary is shown in Fig. \ref{fig:pollock4}.(d).

We also note that the formation of curve liquid filaments has been previously studied due to its importance in prilling processes \cite{Wong2004,PARTRIDGE2005,Decent2009}. Also, the study and prediction of the stability of liquid jets is of great importance for ink-jet printing \cite{Martin2008}. 

\newpage
\section{Techniques}
In this section we discuss some other painting techniques that are not necessarily associated with a particular artists; however, we will use some specific examples by well-known artists.

\subsection{Decalcomania and Saffman-Taylor instabilities}
Decalcomania is a painting technique that is used to create striations, finger-like textures, on a painting. Although its use can be traced back to England in the XVIII$^{th}$ century, to transfer engravings and prints to pottery, the surrealist movement rediscovered it in the 1930s and used it to paint. Interestingly, the technique was considered to be `automatic': the process of creation was supposed to be suppressed of conscious control. In other words, the decalcomania texture appeared without the control of the artist. Oscar Dominguez \cite{Castro1978}, a Spanish surrealist painter, has been credited to first use it in his work. Figure \ref{fig:decal1} shows a painting that displays the technique prominently by contemporary artists Dave Whatt. The texture can be observed all through out the painting but more clearly observed on the lower left on the image. To produce this finger-like features the paint is first deposited on the canvas. For the painting shown in Fig. \ref{fig:decal1}, the artist used acrylic paint. While fresh, the paint is covered with a piece of cloth or paper, which is then ripped off quickly. The stria or fingers appear as a result of this process: the wetting line recedes in a non-uniform manner. Among other notable painters that used this technique we can mention Max Ernst, Remedios Varo and Hans Bellmer.
\begin{figure}[ht!]
    \centering
    \includegraphics[width=0.7\textwidth]{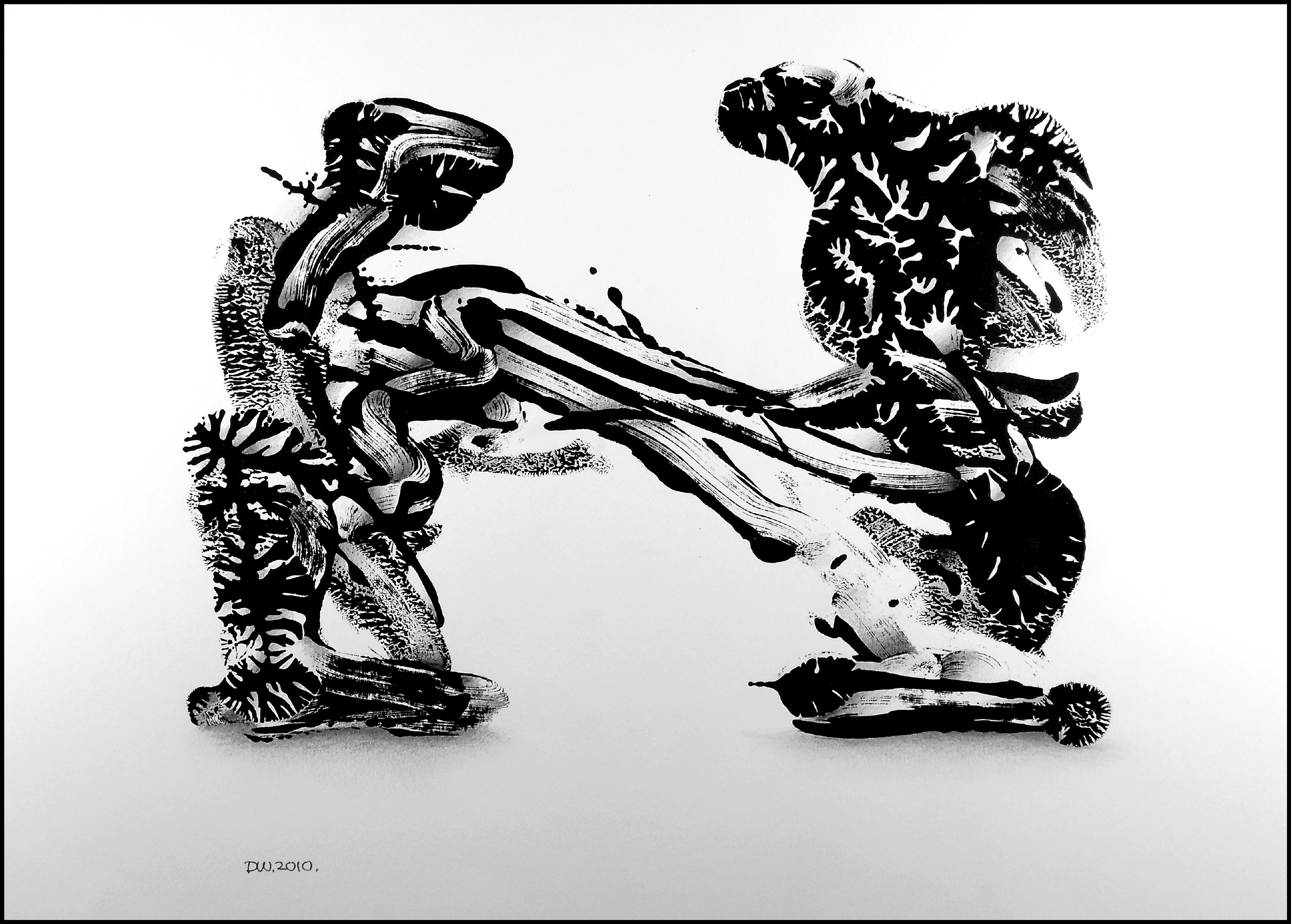}
    \caption{`Untitled', D. Whatt, 2010, reproduced with permission. }
    \label{fig:decal1}
\end{figure}

We speculate that the pattern formation process is the result of the Saffman-Taylor instability \cite{Saffman1958}: the interface between two immiscible fluids becomes unstable if the less viscous fluid advances onto the more viscous one. The unstable undulations grow to form the so-called fingers. This instability has been studied vastly in many contexts, with particular importance for the case of porous media \cite{Homsy1987} due to its implications for the flow in oil reservoirs.  As opposed to the classical Saffman-Taylor instability, the flow in decalcomania is not driven by a pressure gradient; instead, it is driven by mass conservation: the fluid moves due to the `extraction' of fluid in the perpendicular direction of the motion of the contact line. This configuration has been studied by several authors \cite{Nase2011,Dias2013}. The problem has significant importance for the adhesives industry \cite{Pizzi2017}.

To gain some insight into the mechanisms that lead to the formation of decalcomania fingers, we conducted a few controlled experiments. A certain volume of paint (transparent lacquer) was placed between two transparent acrylic plates, separated a certain initial distance. For a given fluid volume, the aspect ratio of the confined fluids is given by $C_o=R_o/h$, where $R_o$ is the initial radius of the confined fluid drop and $h$ is the separation between plates. The top plate is lifted vertically at a constant speed, $U_{lift}$. The interface, or contact line, between the lacquer and the air moves inwards as the separation between the two plates increases in time. The flow is filmed from below with a high speed camera. This configuration is the same as that used by Nase \etal \cite{Nase2011}. Figure \ref{fig:decal2} shows snapshots of the process for different conditions at a certain time, when the fingers have grown fully. Although the configuration of the experiment and the decalcomania technique are not the same, being circular instead of linear, the resemblance of the texture formed by the fingers is noteworthy. 
\begin{figure}[ht!]
    \centering
    \subfigure[\,$C_o=27, Ca= 9$]{\includegraphics[width=0.32\textwidth]{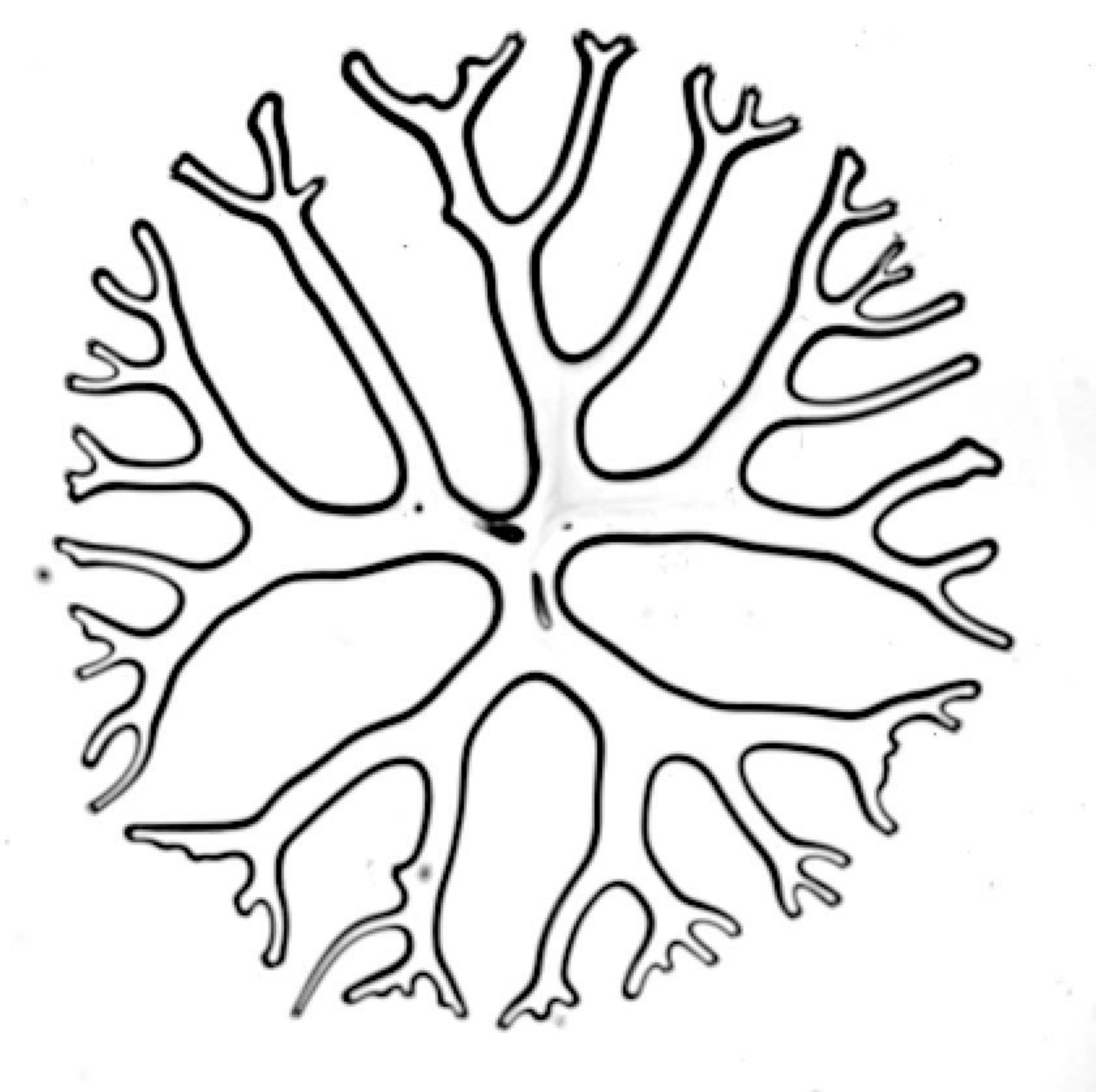}}  \,
    \subfigure[\,$C_o=90, Ca= 9$]{\includegraphics[width=0.32\textwidth]{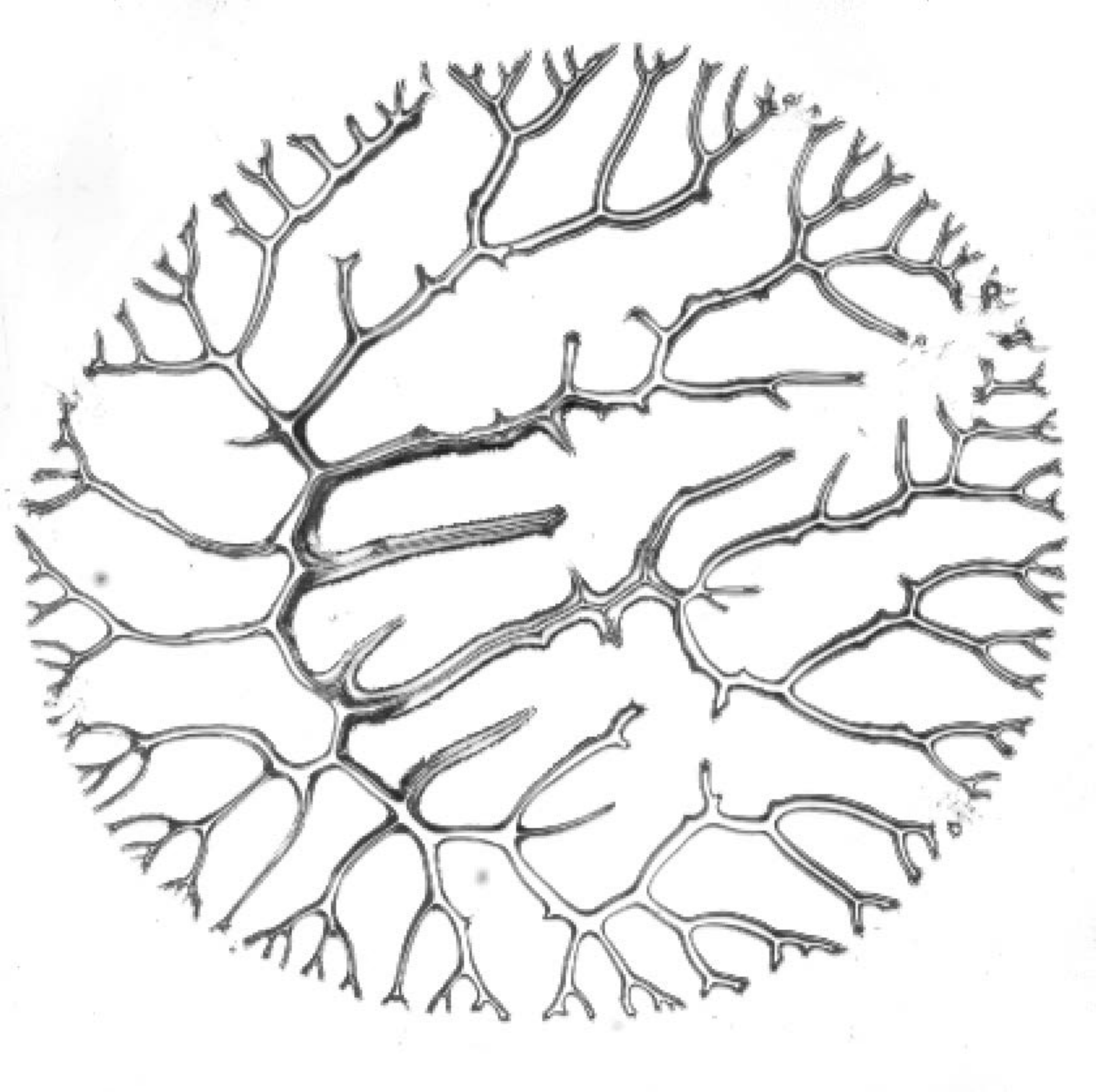}}  \,
    \subfigure[\,$C_o=26, Ca= 48$]{\includegraphics[width=0.32\textwidth]{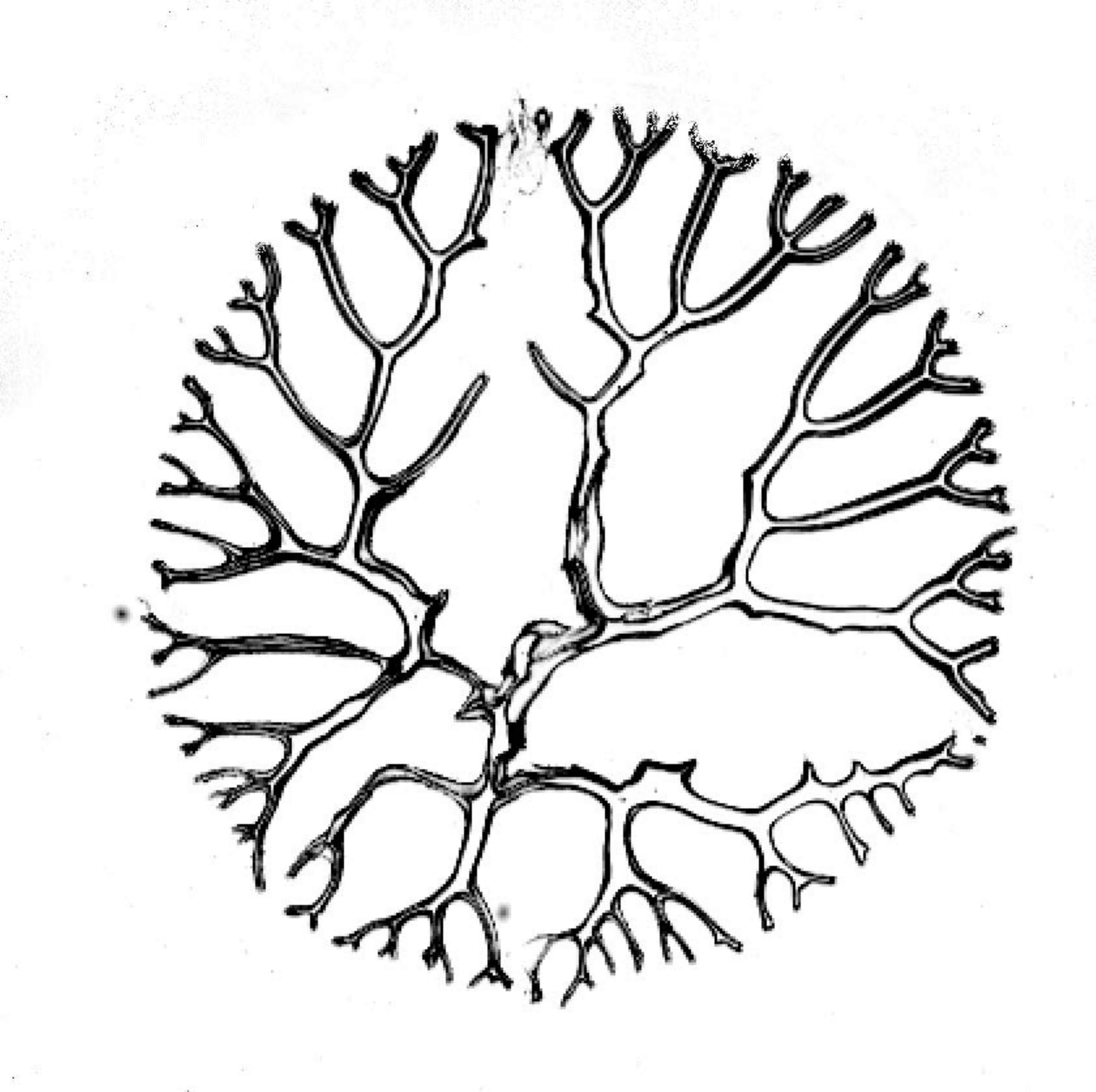}}      
    \caption{Finger formations by the lifting-plate configuration described in the text. In all cases, $t^*\approx 5$.}
    \label{fig:decal2}
\end{figure}

Nase \etal \cite{Nase2011} conducted a linear instability calculation to determine the number of fingers, $N$, at the interface:
\begin{equation}
    N^2 \sim 1+\frac{Ca \, C_o^3}{2(1+t^*)^{9/2}} 
\end{equation}
where $t^*=t U_{lift}/h$ and $Ca$ is the capillary number, $Ca=\mu U_{lift}/\sigma$, where $\mu$ and $\sigma$ are the fluid viscosity and surface tension.  Therefore, the number of fingers decreases in time; larger values of the capillary number or confinement lead to more fingers. The results shown in Fig. \ref{fig:decal2} are in agreement with this estimation and also in good qualitative agreement with \cite{Nase2011}. We find that, for a given time, the number of fingers increases with either the confinement $C_o$ or the value of $Ca$. So, based on these arguments, the technique would not be `automatic' (as originally argued by the surrealistic painters): the number of fingers would change depending on how thin the layer of paint is, how viscous and how fast is the lift off of the covering sheet.

We are currently conducting more experiments considering non Newtonian fluid properties and also different values of the wettability of the substrate. Artist who use decalcomania manipulate the properties of the canvas, by adding wax and other surface coatings to obtain `better' results. We also plan to change the configuration to consider a linear contact line{, which resembles more closely the decalcomania configuration. This configuration has, of course, been studied previously \cite{BenJacob1985}; some authors have argued that such apparently minor changes in configuration may lead to important differences in the appearance of the instability \cite{AlHousseiny2012}. Another factor of possible importance is the use of a flexible lifting surface. If small imperfections greatly affect the growth rate of the unstable fingers \cite{BenJacob1985}, plate flexibility may also play an important role. A complete account of these results will be reported elsewhere.

An interesting aspect of the physical understanding of these painting techniques is that, given a painting, we can attempt to infer the conditions in which the texture was produced. 
Let us consider
the painting 'Untitled' by O. Dominguez, 1937-37, Museum of Modern Art New York (easily
found via a simple web search). It depicts a lion-like image. The line that limits the mane of the
lion is approximately 4 cm and it has approximately 15 large fingers, we can try to infer the speed
of the detachment process if the paint properties are known. Considering that the linear stability calculation from \cite{Nase2011} is valid, $N$ would roughly represent the number of fingers per unit length. Considering  a dimensionless time $t^*=5$ (when only large fingers remain), and a confinement of $C_o=80$ (assuming $h=0.5$ mm), leads to a value of $Ca\approx 210$. Assuming $\mu=0.5$ Pa s and $\sigma=0.04$ N/m, we can infer that $U_{lift}=0.67$ m/s, which is a reasonable and attainable speed for the action of ripping the cover paper over the paint. More or less fingers could, therefore, be produced by changing the speed $U_{lift}$, the initial thickness $h$ or even the paint viscosity $\mu$. A similar exercise can be attempted for other painting techniques, once the physical process is understood.

\subsection{Watercolors and the coffee ring}
Watercolor painting is perhaps the way in which most of us are introduced to art. It is also one of the most ancient painting techniques: it is believed that the cave paintings in Europe were executed with water-based inks. It is also a technique considered difficult to master, since it is hard to correct mistakes and to color large areas uniformly. In general, watercolor painting is conducted by depositing water-based dyes over a porous surface, generally paper. The dyes are a mixture of water, pigments and a binding media (usually gum arabic). The paper is porous and absorbent, made with wood pulp and cotton fibers, but is usually covered with gelatine. Artists normally apply the dye onto the dry paper but in many instances, the surface is pre-wetted before applying the paint. 
\begin{figure}[ht!]
    \centering
    \includegraphics[width=0.7\textwidth]{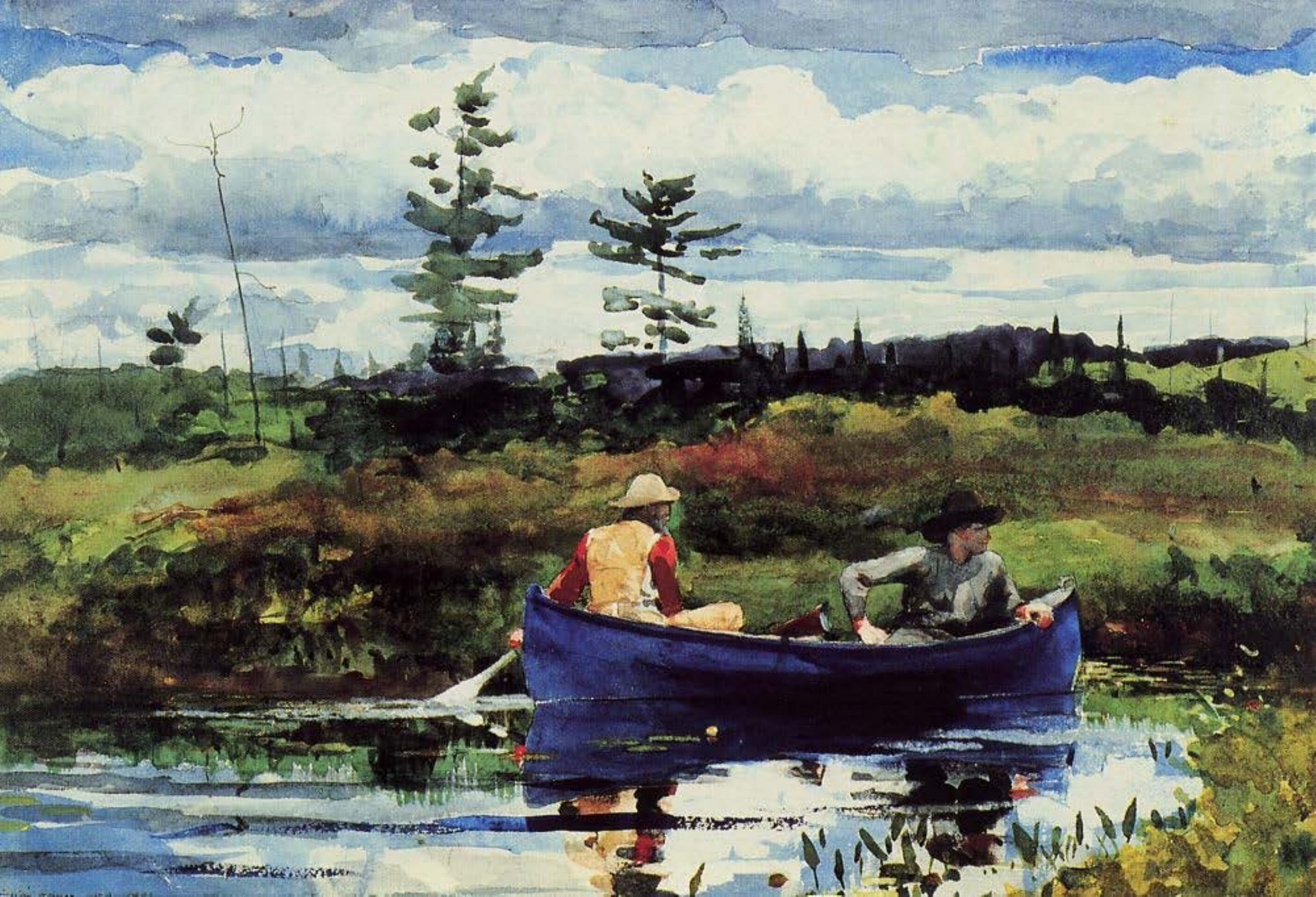}
    \caption{The Blue Boat, W. Homer, 1892, Museum of Fine Arts, Boston. Reproduced under public domain permission. }
    \label{fig:water1}
\end{figure}

While the subject matter is, of course, dependent on the style and skill of the painter, the formation of textures is dominated by the flow that occurs while applying the paint and, particularly, during the drying process. Figure \ref{fig:water1} show an example of  watercolor painting by Winslow Homer, an American landscape painter, considered one of the most distinguished American painters in XIX$^{th}$ century. The composition of the painting contains regions where the pigment is uniform or mixed, with color intensity gradients and areas with hardly any colors. 

It turns out that, for water-based inks deposited on dry paper, it is actually hard to produce regions of uniform color intensity. In most cases, the pigment particles are small; hence, they follow the motion of the fluid. A sessile evaporating drop of an aqueous liquid with small particles, has been shown to form the so-called `coffee ring' stain. This effect has been vastly studied \cite{Deegan1997} and there is a clear understanding of its underlying physics. In essence the fluid motion inside the drop occurring during the evaporation process, under certain conditions, leads to the preferential deposit of the particles at the rim of the drop. 
\begin{figure}[ht!]
    \centering
    \includegraphics[width=0.6\textwidth]{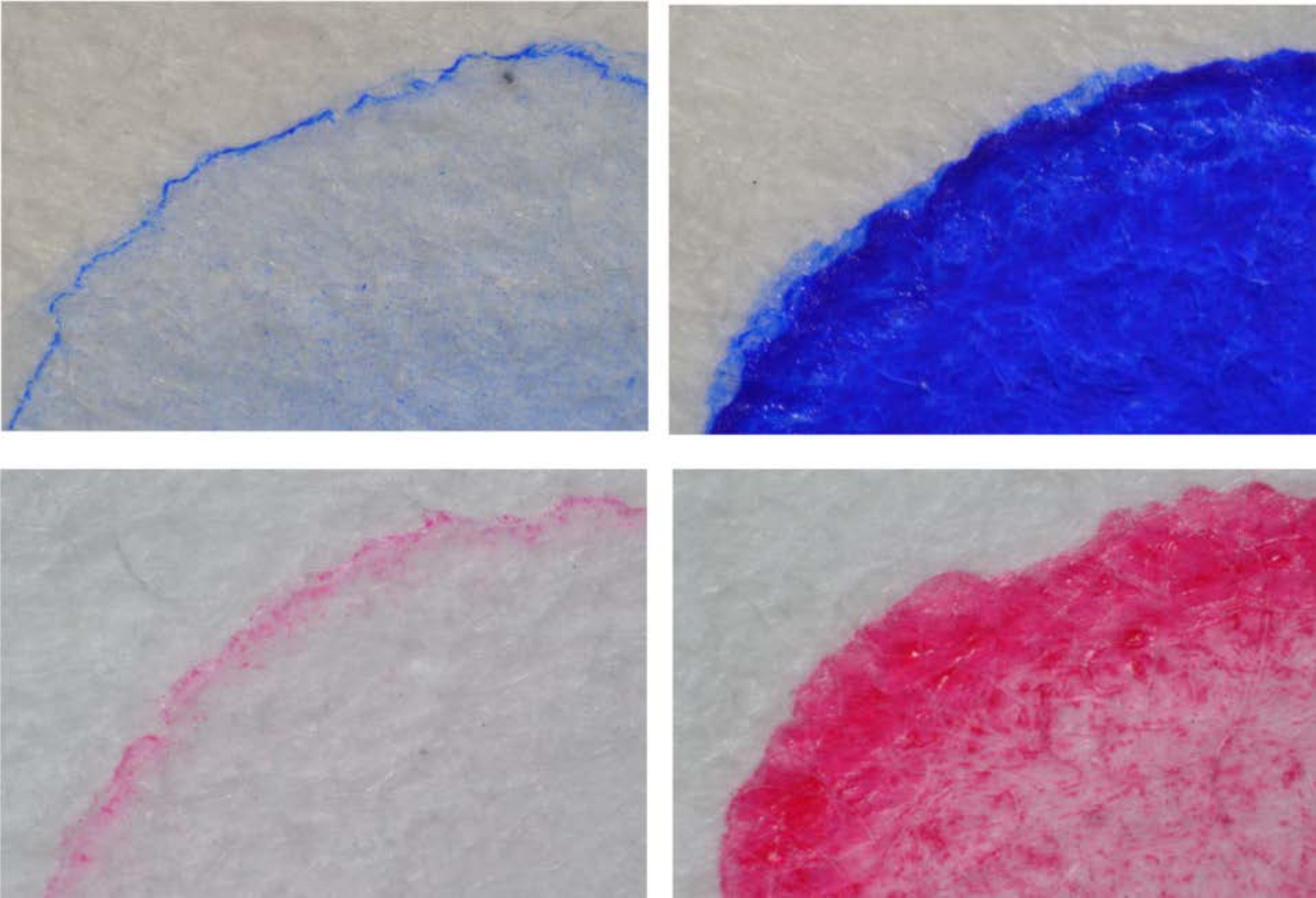}
    \caption{Stains of watercolor dyes applied to dry paper. On the top  and bottom rows, blue ultramarine and genuine rose dyes are shown, respectively. On the left and right columns, tests with small and large pigments concentrations are shown. In both cases, a 50 $\mu$l drop was deposited on dry paper and let evaporate under standard laboratory conditions.}
    \label{fig:water2}
\end{figure}

In order to understand the conditions for which the coffee ring stain appears  for the case of watercolors, we conducted a series of experiments considering different inks deposited on dry paper. A detailed account of these tests can be found in \cite{Gonzalez2019}. Figure \ref{fig:water2} shows examples of the type of stains left by two different pigments and  pigment concentrations. For small pigment concentrations, the coffee stain is clearly observed (left side of Fig. \ref{fig:water2}); for large concentrations (right side of Fig. \ref{fig:water2}), the blue pigment produces a nearly uniform intensity stain but the pink one still shows a coffee ring but with a thicker rim. The difference between the two pigments, in addition to the color, is the particle size. For the case of blue, the particles are large, so they do not necessarily follow the fluid flow during the during process. Hence, when the concentration is large, the particles sediment leading to a nearly uniform color intensity stain. On the other hand, the pink particles are small; therefore, the are always transported by the fluid flow resulting in their accumulation at the rim of the stain. Note that in this case, the sessile drops remained pinned during the drying process. These results are in good agreement with \cite{Shen2010}, where the appearance and size of the coffee stain rim is addressed.

Water colorists, therefore, learn empirically how to avoid or promote this effect. Many other textures can also be produced by changing the the wettability condition of the paper, that is by applying water before painting. Figure \ref{fig:water3} shows an example, using the blue dye, of the textures that can be produced by changing the amount of water in the paper before depositing the dye. Note that in the three cases shown, the same pigment concentration is considered. In this case, in addition to the evaporation-induced pattern formation, the dye permeates through the paper changing the pigment deposition dynamics before evaporation. A full account of the effect of paper wetness on the pattern formation for watercolors can be found in \cite{Gonzalez2019}.
\begin{figure}[ht!]
\includegraphics[width=0.7\textwidth]{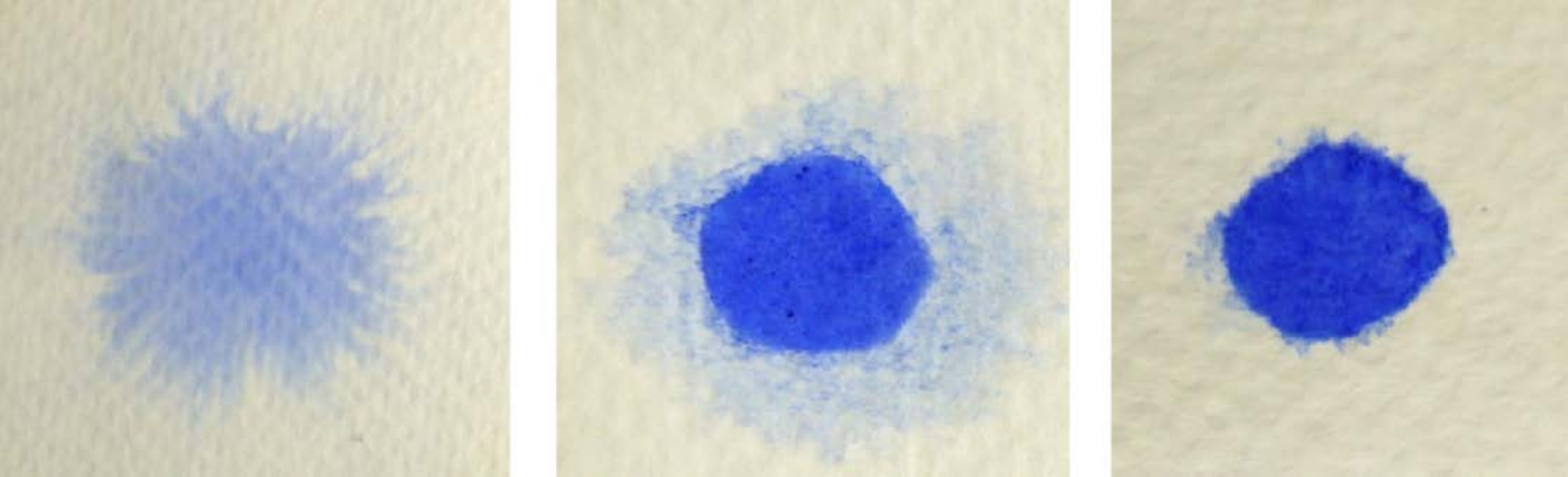}
    \caption{Stains of watercolor dyes applied to wet paper. The water content in the three cases is the same but the distribution on the surface is different. The results are for ultramarine blue drops of 50 $\mu$l, dried over laboratory conditions.}
    \label{fig:water3}
\end{figure}

As with the previous examples, the study of drying sessile droplets  also has  important practical implications. The shape and texture of the patterns left by evaporated drops have been used as health indicators \cite{Sefiane2010}, to quantify bacterial motility \cite{Nellimoottil2007}, to determine the calorimetric properties of  membranes \cite{Gonzalez2017b} and to authenticate consumable beverages \cite{Gonzalez2017}, among many others uses.




\section{Conclusions and outlook}

In this paper, an analysis of a few artistic painting techniques has been presented. While painting, either industrial or artistic, share the same mechanical principles, the later has not been studied extensively. Understanding the processes which dictate the formation of textures and patterns in artistic painting is of importance, first, from a fundamental point of view; paraphrasing Feynman: `What I do not understand, I cannot create' (quote taken from his blackboard at the time of his death). We interpret this quote as the need to understand the underlying physics of a process in order to be able to control it and reproduce it. This is exactly our  view to study the mechanics of artistic painting. The process is bound by physical constrains, the textures are dictated by the flow induced during the painting action. Artists are, therefore, fluid mechanicians but they do not know it. Second, understanding the mechanics of how a painting was executed could be of value to place it in time. In other words, fluid mechanics could be used as a tool in art history or for the authentication of works of art. Producing certain compositions or paintings would be constrained by material properties and techniques that can be located to a specific historical period. Similarly, the restoration and preservation of art works could be conducted more effectively if the production processes are understood. And, finally, studying flows in artistic painting could provide inspiration to address issues in other areas of fluid mechanics. For instance, to understand salt domes, one could consider the configuration used in Siqueiros accidental painting technique, under laboratory and controlled conditions, naturally, considering appropriate scaling laws.

We end with a comment about an unexpected by-product of these investigations. Since we wrote our first paper on the subject \cite{Zetina2015}, we have had the opportunity to have discussions with practising artists. They are intrigued by such research projects and are fully engaged in trying to understand the main results. Figure \ref{fig:artists} shows three examples of art works that have been produced as a result of conversation with artists. Figure \ref{fig:artists}(a) shows a painting by Julie Underriner, who saw our video in You Tube about the accidental painting technique (presented in the Gallery of Fluid Motion in 2013). After a few email exchanges, she was able to reproduce the classical texture of the technique by inducing the Rayleigh-Taylor instability. Her piece is, of course, more colorful than our experiments but both hold the same physical principles. Figure \ref{fig:artists}(b) shows a recent painting by Arturo Buitron, who visited our laboratory. He heard a presentation about the flying viscous catenaries of Pollock and saw the experiment working. Shortly after, he reproduced the technique in his studio and used the flying catenaries to draw the central theme of his piece. Lastly, Fig. \ref{fig:artists}(c) shows a watercolor painting by Octavio Moctezuma. He also visited our laboratory, but we invited him to paint `in house' to document the watercolor technique in detail. Resulting from his visits, we detected several interesting aspects of watercolor, including the coffee ring effect and the importance of paper humidity. We conducted an in-depth investigation and wrote a paper together \cite{Gonzalez2019}. Therefore, we can further argue that the results of these studies can also be of interest to other communities: art students and practicing artists. Understanding the flow physics can greatly reduce tedious repetition and experimentation and can lead to new ways to paint.
\begin{figure}
    \centering
    \subfigure[]{\includegraphics[width=0.32\textwidth]{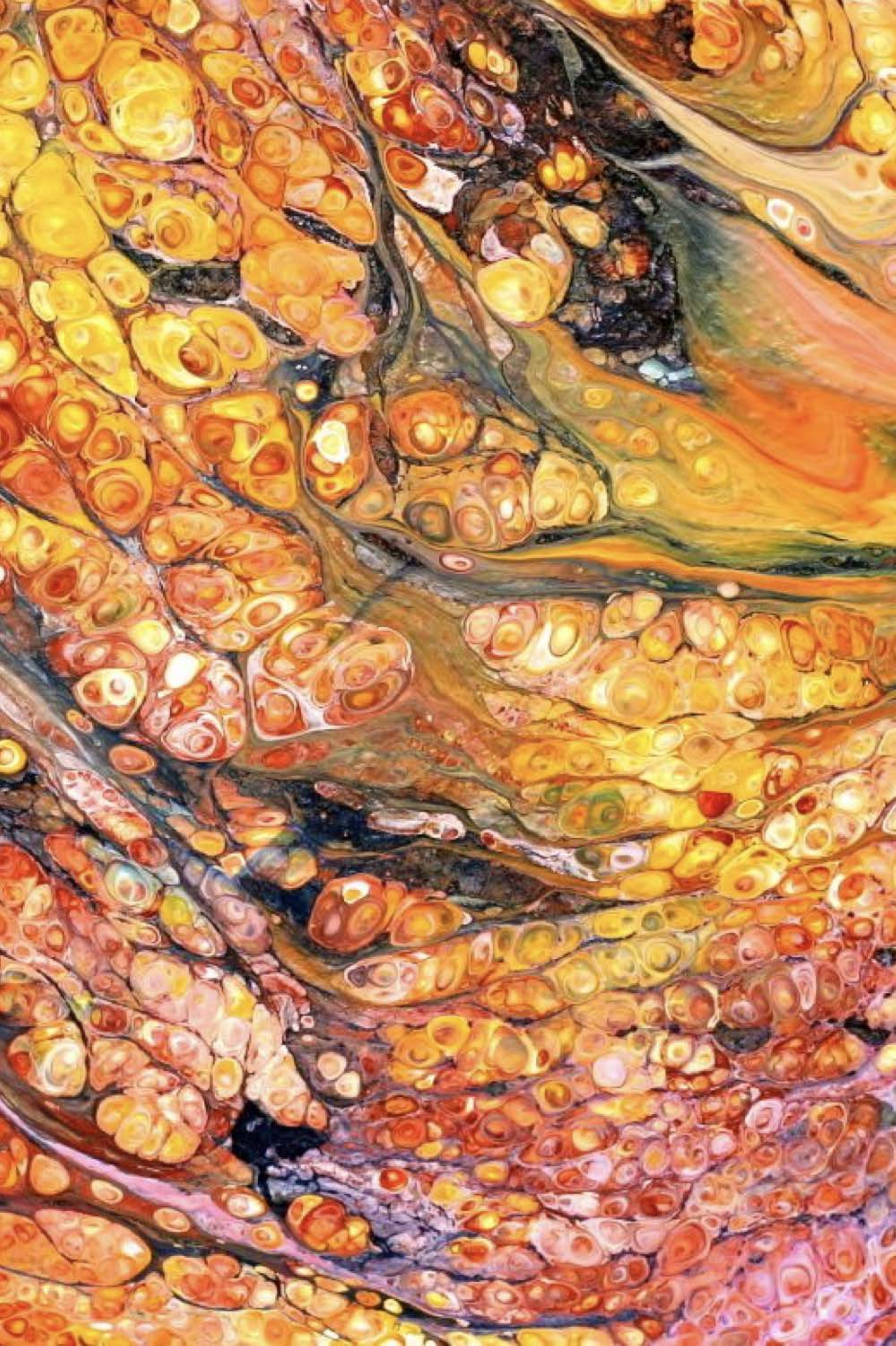}}\,
    \subfigure[]{\includegraphics[width=0.32\textwidth]{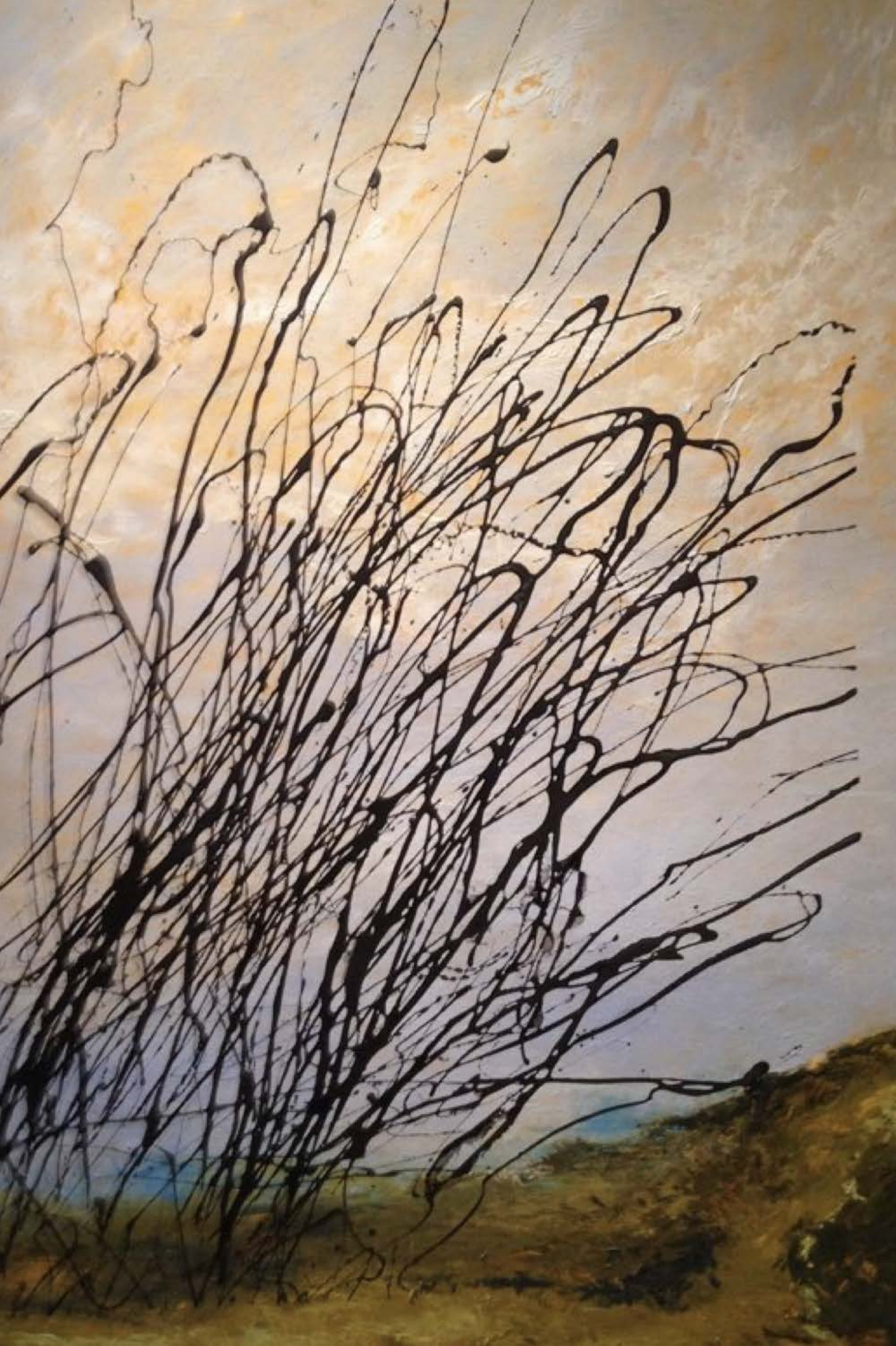}}\,
    \subfigure[]{\includegraphics[width=0.32\textwidth]{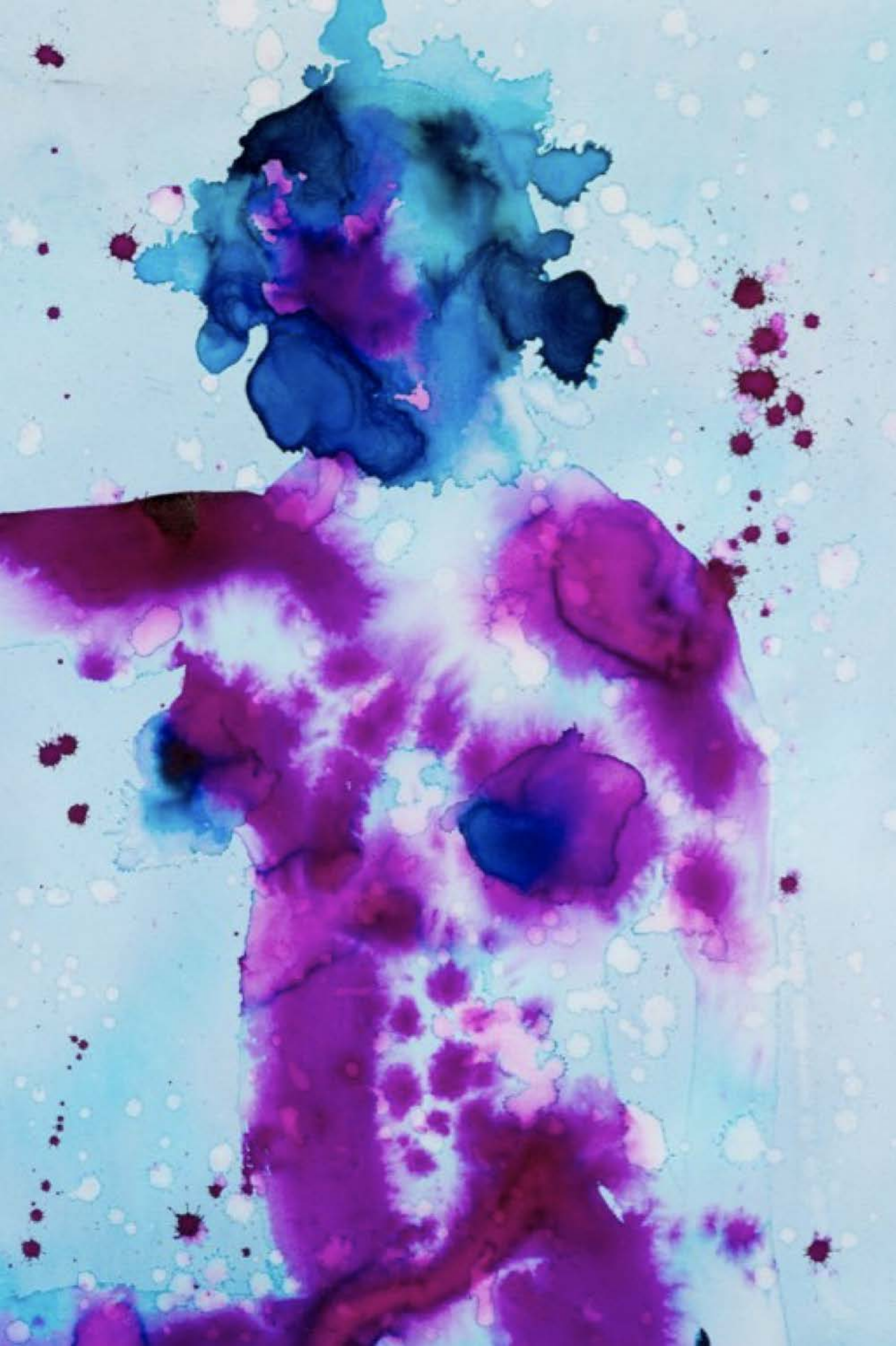}} 
    \caption{Three painting created by artist who  interacted with this investigation. (a) `Gold whirl' (zoom), J. Underriner, 2014; (b) `Land art' (zoom), A. Buitr\'on, 2018; (c) `Fetiche y tab\'u' (zoom), O. Moctezuma, 2019. All images reproduced with permission from the authors.}
    \label{fig:artists}
\end{figure}

\begin{acknowledgments}
The author would like to thank B. Palacios, E. de la Calleja, J. Gonzalez-Gutierrez, I. Farias and A. Rosario for their invaluable help to conduct experiments, and S. Zetina and O. Moctezuma for inspiring discussions. I am grateful for the careful reading and suggestions by Y. Su. The financial support of DGAPA-PAPIIT-UNAM (grant number IN108016) and ACT-FONCA (grant number 04S.04.IN.ACT.038.18) is greatly acknowledged. 
\end{acknowledgments}

\providecommand{\noopsort}[1]{}\providecommand{\singleletter}[1]{#1}%

\end{document}